\documentclass[fleqn,usenatbib]{mnras}
\usepackage{newtxtext,newtxmath}

\usepackage[T1]{fontenc}

\DeclareRobustCommand{\VAN}[3]{#2}
\let\VANthebibliography\thebibliography
\def\thebibliography{\DeclareRobustCommand{\VAN}[3]{##3}\VANthebibliography}

\usepackage{graphicx}	% Including figure files
\usepackage{amsmath}	% Advanced maths commands
\usepackage[separate-uncertainty=true]{siunitx}
\usepackage{textcomp}
\usepackage{CJK}
\usepackage{gensymb}

\newcommand{\obj}{{WD1032+011}}
\newcommand{\obja}{{WD1032+011A}}
\newcommand{\objb}{{WD1032+011B}}
\newcommand{\objab}{{WD1032+011AB}}

\title[The only inflated eclipsing WD-BD binary]{The only inflated brown dwarf in an eclipsing white dwarf--brown dwarf binary: WD1032+011B}

\author[J. R. French et al.]{Jenni R. French,$^{1}$\thanks{E-mail: jf328@leicester.ac.uk (JRF)}
Sarah L. Casewell,$^{1}$
Rachael C. Amaro,$^{2}$
Joshua D. Lothringer,$^{3}$
L. C. Mayorga,$^{4}$
\newauthor
Stuart P. Littlefair,$^{5}$
Ben W. P. Lew,$^{6}$
Yifan Zhou,$^{7}$
Daniel Apai,$^{2,8}$
Mark S. Marley,$^{8}$
Vivien Parmentier,$^{9,10}$
\newauthor
Xianyu Tan,$^{11}$
\\
$^{1}$Centre for Exoplanet Research, School of Physics and Astronomy, University of Leicester, University Road, Leicester, LE1 7RH, United Kingdom\\
$^{2}$Department of Astronomy and Steward Observatory, University of Arizona, 933 North Cherry Avenue, Tuscon, AZ 85721, USA\\
$^{3}$Space Telescope Science Institute, 3700 San Martin Drive, Baltimore, MD 21218, USA\\
$^{4}$Johns Hopkins University Applied Physics Laboratory, 11100 Johns Hopkins Road, Laurel, MD 20723, USA\\
$^{5}$Department of Physics and Astronomy, University of Sheffield, Sheffield, S3 7RH, United Kingdom\\
$^{6}$Bay Area Environmental Research Institute and NASA Ames Research Center, Moffett Field, CA 94035, USA\\
$^{7}$Department of Astronomy, University of Virginia, 530 McCormick Road, Charlottesville, VA 22904, USA\\
$^{8}$Lunar and Planetary Laboratory, University of Arizona, 1629 East University Boulevard, Tucson, AZ 85721, USA\\
$^{9}$Atmospheric, Oceanic and Planetary Physics, Department of Physics, University of Oxford, Parks Road, Oxford, OX1 3PU, United Kingdom\\
$^{10}$Universit\'{e} C\^{o}te d'Azur, Observatoire de la C\^{o}te d'Azur, CNRS, Laboratoire Lagrange, France\\
$^{11}$Tsung-Dao Lee Institute \& School of Physics and Astronomy, Shanghai Jiao Tong University, Shanghai 201210, China\\
}

\date{Accepted XXX. Received YYY; in original form ZZZ}

\pubyear{2024}

\begin{document}
\label{firstpage}
\pagerange{\pageref{firstpage}--\pageref{lastpage}}
\maketitle

\begin{abstract}
Due to their short orbital periods and relatively high flux ratios, irradiated brown dwarfs in binaries with white dwarfs offer better opportunities to study irradiated atmospheres than hot Jupiters, which have lower planet-to-star flux ratios. \obj{} is an eclipsing, tidally locked white dwarf--brown dwarf binary with a 9950~K white dwarf orbited by a 69.7~M$_{\text{Jup}}$ brown dwarf in a 0.09~day orbit. We present time-resolved Hubble Space Telescope Wide Field Camera 3 spectrophotometric data of \obj{}. We isolate the phase-dependent spectra of \objb{}, finding a 210~K difference in brightness temperature between the dayside and nightside. The spectral type of the brown dwarf is identified as L1 peculiar, with atmospheric retrievals and comparison to field brown dwarfs showing evidence for a cloud-free atmosphere. The retrieved temperature of the dayside is 1748$^{+66}_{-67}$~K, with a nightside temperature of 1555$^{+76}_{-62}$~K, showing an irradiation-driven temperature contrast coupled with inefficient heat redistribution from the dayside to the nightside. The brown dwarf radius is inflated, likely due to the constant irradiation from the white dwarf, making it the only known inflated brown dwarf in an eclipsing white dwarf--brown dwarf binary.

\end{abstract}

\begin{keywords}
white dwarfs - brown dwarfs - binaries
\end{keywords}

%%%%%%%%%%%%%%%%%%%%%%%%%%%%%%%%%%%%%%%%%%%%%%%%%%

%%%%%%%%%%%%%%%%% BODY OF PAPER %%%%%%%%%%%%%%%%%%

\section{Introduction}

There is an observed scarcity of brown dwarfs orbiting main sequence stars within 3~AU, which is termed the `brown dwarf desert' \citep{BDdesert2, BDdesert1}. An analysis of the brown dwarf desert by \cite{BDdesert2} found that in a sample of 514 stars with a companion object within 10~AU, only 2 of these were brown dwarfs. As the main sequence star in these systems evolves along the giant or asymptotic giant branch, it expands and fills out its Roche Lobe, leading to Roche Lobe overflow \citep{RLOF}. Brown dwarfs cannot accept the incoming material, and the red giant's envelope engulfs the brown dwarf. A brief phase of binary evolution then occurs in a common envelope, where the brown dwarf does not have sufficient mass to force the envelope to co-rotate \citep{CEE}. Friction causes the binary orbit to decay as the companion loses orbital angular momentum to the envelope which is then ejected \citep{evolution}. The resultant system is a close, post-common envelope white dwarf--brown dwarf binary. \newline
\indent Since the brown dwarf must survive being engulfed by the white dwarf's progenitor, these systems are rare, and only $\sim$0.1--0.5\% of white dwarfs are predicted to have a brown dwarf companion \citep{bdrates, Steele, Rebassa}. There are several all-sky surveys that have generated white dwarf catalogues \citep{WD1, WD2}, however there are currently only 11 known close, detached white dwarf--brown dwarf binaries: GD1400 (WD+L6, P=9.98~hrs; \citealt{GD1}), WD0137-349 (WD+L6--L8, P=116~min; \citealt{0137}), WD0837+185 (WD+T8, P=4.2~hrs; \citealt{0837}), 
NLTT5306 (WD+L4--L7, P=101.88~min;
\citealt{NLTT}),
SDSS J141126.20+200911.1 (WD+T5, P=121.73~min; \citealt{J1411}),
SDSS J155720.77+091624.6 (WD+L3--L5, P=2.27~hrs; \citealt{J155}), SDSS J1205-0242 (WD+L0, P=71.2~min; \citealt{J1205}), SDSS J1231+0041 (WD+M/L, P=72.5~min; \citealt{J1205}), EPIC212235321 (WD+L3, P=68.2~min; \citealt{epic}), WD1032+011 (WD+L4--L6, P=2.20~hrs; \citealt{wd1032}) and ZTF~J0038+2030 (WD+BD, P=10.4~hrs; \citealt{ztf}). Additionally, recent analysis of eclipsing white dwarfs in the Zwicky Transient Facility Survey have identified several candidate eclipsing white dwarf--brown dwarf binaries \citep{ztf1, ztf2, ztf3}. The brown dwarfs in these systems are likely tidally locked, as hot Jupiters are, and the irradiation of the brown dwarf results in temperature differences between the `day' and `night' side of up to 500~K \citep{irrad}. Eventually, the brown dwarf companion will lose sufficient orbital angular momentum such that mass transfer will begin, forming a Cataclysmic Variable with a sub-stellar donor \citep{CV, CV2}.

Despite their rarity, close white dwarf--brown dwarf binaries provide insights into sub-stellar object survival in common envelope evolution and an opportunity to study models of irradiated atmospheres \citep{irrad, irrad2}. Irradiated white dwarf--brown dwarf binaries bridge the gap between non-irradiated brown dwarfs and irradiated hot Jupiters. Some ultrahot brown dwarfs, such as EPIC2122B, have temperatures between that of the hottest hot Jupiter (KELT-9b, $T_{\text{eq}} = 4050$~K, \citealt{kelt9b}), and next hottest (TOI-2109b, $T_{\text{eq}} = 3646$~K \citealt{TOI-2109}; WASP-33b, $T_{\text{eq}} = 2800$~K, \citealt{wasp33b}), offering more objects to study in sparsely populated areas of the parameter space encompassing hot Jupiters. Hot Jupiters are a class of exoplanets which have masses $M \geq 0.25~\text{M}_{\text{Jup}}$, and orbit their host stars with periods of 10~days or less \citep{HJparams}. They comprise some of the most well-studied planetary-mass objects, with over 300 discovered to date \citep{exoarchive}. They are tidally locked to their host stars, and receive significant irradiation on one hemisphere as a result \citep{tidallylocked}. It is difficult to characterise the atmospheres of hot Jupiters well due to the poor flux ratio between the planetary signal and the host star. However, the atmospheres of brown dwarfs have been well studied, and where the host star is a white dwarf, the contaminant flux in the infrared is minimal, leading to higher planet-to-star flux ratios. Irradiated brown dwarfs thus provide excellent proxies with which to study irradiated atmospheres and hot Jupiters.

Recent spectroscopic studies of hot Jupiters have revealed temperature differences between the dayside and nightside atmospheres on the order of a few hundred Kelvin (e.g., \citealt{HJ1, atmosstruc}) that can be as large as $\sim$1000~K \citep{HJ1000K}. These large temperature differences influence a range of atmospheric dynamics including atmospheric structure \citep{atmosstruc}, jets \citep{jets}, and turbulence \citep{turbulence}. Heat transport between the dayside and nightside in hot Jupiters is mainly enabled by the presence of equatorial jets, and several recent works have investigated how parameters such as atmospheric composition and rotation rate influence this heat transport (e.g., \citealt{atcomp, hjatmos}).

3D circulation models which have been applied to irradiated brown dwarfs atmospheres show that the dayside hot spot does not undergo eastward-shifting, unlike those seen in hot Jupiters \citep{Lee2020, Wong2016}. The equatorial jets that are vital for heat redistribution are shrunk due to the fast rotation rates seen in brown dwarfs, thus suppressing the heat transfer from the dayside to the nightside \citep{TanShow2020}. Radiative transfer and chemical equilibrium modelling of the most highly irradiated brown dwarfs has found that molecules at the upper atmosphere and photosphere of irradiated brown dwarfs are effectively dissociated, resulting in weak molecular absorption in the dayside atmosphere. Additionally, the atomic emission lines seen in ultrahot irradiated brown dwarf atmospheres could be due to a thermal inversion that is caused by the strong ultraviolet heating from the white dwarf \citep{lothcase2020}. These temperature inversions and ionised hydrogen atoms arise because the upper atmosphere of the brown dwarf can absorb the short wavelength irradiation from the white dwarf more easily than deeper layers of the atmosphere \citep{yifan}.

Many hot Jupiters have been found to be inflated, that is to say that their radius is larger than what is predicted by planetary structure models considering their mass and equilibrium temperature (e.g., \citealt{inflated1, fortney, inflated2, inflated3}). If a planet's equilibrium temperature, $T_\text{eq}$, is greater than 1200~K, the difference between its model-predicted radius and the observed one increases as $R \propto T_{\text{eq}}^{1.4}$ \citep{inflatedeq}. Thus, the greater the irradiation, the more inflated the radius of the planet. Irradiated brown dwarfs can also exhibit inflated radii similar to those seen in hot Jupiters, however they do not follow the same linear trend with irradiation flux (e.g., \citealt{J1205, casewell2020}).

Many of the well-studied white dwarf--brown dwarf binary systems are non-eclipsing, and physical parameters of the brown dwarf can therefore only be estimated from evolutionary models. If the orbital plane of white dwarf--brown dwarf binary is along our line of sight, then the brown dwarf will eclipse the white dwarf, and vice versa, during an orbit. White dwarfs have typical radii 10$\times$ smaller than brown dwarfs, so the brown dwarf will completely occult the white dwarf as it passes in front of it. Since the brown dwarf is tidally locked, its nightside will be observed as it eclipses the white dwarf. Therefore, in eclipsing white dwarf--brown dwarf binaries, any spectra taken whilst the brown dwarf eclipses the white dwarf will be lone spectra of the brown dwarf, uncontaminated by any white dwarf flux. The dayside and nightside spectra of the brown dwarf can then be robustly extracted to investigate the atmospheric dynamics of the brown dwarf (e.g, \citealt{ben}). Eclipsing close white dwarf--brown dwarf binaries thus make important benchmark systems which yield insights into irradiated brown dwarfs and exoplanet atmospheres.

To better understand the effect of irradiation on brown dwarfs and how this influences their atmospheric composition, we present Hubble Space Telescope Wide Field Camera 3 observations of \obj, an eclipsing white dwarf--brown dwarf binary system. Our new observations allow the white dwarf and brown dwarf components to be separated from the combined spectroscopic observations. In Section \ref{sec:target} we discuss the target; in Section \ref{sec:observations} we describe our observations, and our data reduction in described in Section \ref{sec:reduction}. In Section \ref{subsec:lightcurve} we generate and analyse the broadband lightcurve; in Section \ref{subsec:spectra} we present the phase-dependent spectra of the brown dwarf and compare these to models, and Section \ref{subsec:brighttemp} investigates the brightness temperature of our spectra. In Section \ref{sec:fieldbds} we compare to field brown dwarfs; in Section \ref{subsec:nonirrbdmodels} we compare to non-irradiated brown dwarf models; in Section \ref{subsec:forwardmodels} we run forward models, and in Section \ref{subsec:irrbdmodels} we run atmospheric retrievals considering irradiated brown dwarf models. We discuss our results in Section \ref{sec:discussion}.

\section{WD1032+011}
\label{sec:target}
\obj{} was first identified as a DA white dwarf by \cite{vennes}, and \cite{eisenstein} measured an effective temperature of $T_{\text{eff}} = 9904~\pm~109$~K and a surface gravity of log~$g = 8.13~\pm~0.15$ using the Sloan Digital Sky Survey (SDSS). \cite{Steele} found an infrared excess in the UKIRT Infrared Deep Sky Survey (UKIDSS) photometry. They suggested that this was due to an unresolved companion with a spectral type of L$5~\pm~1$ and a mass of $M = 55~\pm~4$~M$_{\text{Jup}}$ orbiting within 150~AU. 

\cite{wd1032} used K2 photometry of \obj{} spanning across $\approx$81~days using long cadence mode, and found a most likely period of $\approx$2.2~hours using a Lomb-Scargle periodogram. Spectroscopy was taken using the Gemini Multi-Object Spectrograph covering a wavelength range of 4600-6900~\AA. After producing trailed spectra centred on the H$\beta$ line, they found a clear oscillation across a full orbit, corroborating the companion detection (Figure 2 in \citealt{wd1032}). They calculated a radial velocity of $\gamma = 122.1 \pm 1.9$~km~s$^{-1}$, and following the same kinematic analysis in \citealt{uvwdisc} determine that \obj{} is a likely member of the thick disc. They thus estimate the age as $>$ 5~Gyr.  

They also observed \obj{} with the Gemini Near-Infrared Spectrograph (GNIRS) across the entire spectrum of 0.8--2.5$\upmu$m with the \SI{1}{\arcsecond} slit. They compared their GNIRS spectrum to composite DA white dwarf + brown dwarf models from the SpeX prism library \citep{spexlib} and determined that a companion spectral type of L5 is most likely, in concurrence \cite{Steele}. Additionally, they compared the UKIDSS magnitudes of \obj{} to the absolute magnitudes of L3--L6 field brown dwarfs from \cite{dupuyliu}, which also indicate a companion consistent with spectral types L4--L6. 

A Spectral Energy Distribution (SED) fit was performed using data from SDSS, GALEX and UKIDSS to estimate a white dwarf mass and determine the effective temperature. Three individual eclipses of \obj{} were observed using ULTRACAM, which simultaneously observes in three different filters. The system is eclipsing and the inclination was constrained to $87.5 \pm 1.4$\degree, but no evidence of a secondary eclipse was found. \cite{wd1032} normalised the ULTRACAM lightcurves and used an affine-invariant Monte Carlo Markov Chain (MCMC) sampler to determine masses and radii for both the white dwarf and the brown dwarf. Table \ref{table:wd1032params} lists the key parameters for \obj, which is the only eclipsing white dwarf-brown dwarf binary in which the brown dwarf is thought to be inflated.

\begin{table}
	\centering
	\caption{System parameters for the \obj{} binary system. Values are reproduced from Table 4 in \protect\cite{wd1032}. Equilibrium temperature is calculated here assuming a Bond albedo of zero. The numbers denoted in brackets represent the uncertainties for the period and ephemeris, which apply to the last two decimal places.}
	\label{table:wd1032params}
	\begin{tabular}{lc} % four columns, alignment for each
		\hline
		Parameter & Value\\
		\hline \\
        White Dwarf $T_{\text{eff}}$ (K) & $9950 \pm 150$\\ \\
        White Dwarf log $g$ & $7.65 \pm 0.13$\\ \\
        White Dwarf Cooling Age (Gyr) & $0.455 \pm 0.080$\\ \\
        Period (days) & $0.09155899610(45)$ \\ \\
        Ephemeris (BMJD) & $58381.2439008(10)$\\ \\
        White Dwarf Radius (R$_{\odot}$) & $0.0147 \pm 0.0013$\\ \\
        White Dwarf Mass (M$_{\odot}$) & $0.4052 \pm 0.0500$\\ \\
        Brown Dwarf Radius (R$_{\odot}$) & $0.1052 \pm 0.0101$\\ \\
        Brown Dwarf Radius (R$_{\text{Jup}}$) & $1.024 \pm 0.098$\\ \\
        Brown Dwarf Mass (M$_{\text{Jup}}$) & $69.7 \pm 6.4$\\ \\
        Brown Dwarf $T_{\text{eq}}$ (K) &$1030 \pm 50$ \\ \\
        Orbital Separation (AU) & $0.00319 \pm 0.00011$ \\ \\ 
        Inclination ($\degree$) &$87.5 \pm 1.4$ \\ \\
		\hline
	\end{tabular}
\end{table}

\section{Observations}
\label{sec:observations}

We observed \obj{} with the Hubble Space Telescope (HST) Wide Field Camera, using the WFC3/IR/G141 grism. We observed across 6 consecutive orbits of HST on the 15th May 2022, as a part of program GO-16754 (PI: S. L. Casewell). In order to perform wavelength calibration, a direct image was taken at the beginning of each orbit using the F127M filter, with the GRISM256 aperture and a subarray setup of 256$\times$256. After these direct images, 8 spectroscopic exposures were taken for each orbit. These spectra were taken in \textit{staring mode} each with an exposure time of 313~s using the G141 grism, the GRISM256 aperture, and the same subarray setup as the direct images. This observing sequence has already been successfully conducted on over a dozen isolated brown dwarfs \citep[e.g.,][]{apai, Lew2016} and close white dwarf--brown dwarf binaries, offering spectra with a good signal-to-noise \citep{ben, yifan, rachael}. Our observations offer full phase coverage of \obj{} across the 6 HST orbits, allowing us to study any phase-dependent changes in its spectra.

\section{Data Reduction}
\label{sec:reduction}

We downloaded our \texttt{flt} file data from the Mikulski Archive for Space Telescopes \citep{mast} after they had been preprocessed by the \texttt{CalWFC3} pipeline \citep{calwfc3}. The \texttt{CalWFC3} pipeline corrects for bias and dark current as well as flagging bad pixels and flat-fielding the image\footnote{\url{https://hst-docs.stsci.edu/wfc3dhb}}.

To extract the spectral data from the \texttt{flt} files, we utilised the established pipeline used by \cite{rachael}, which is the latest iteration of a pipeline developed and adapted by \cite{buenzli} and \cite{apai}. This pipeline is an amalgamation of the \texttt{aXe} software designed for reducing HST WFC3 data \citep{axe} and a custom program written in \texttt{Python}. This pipeline has been shown to be successful in reducing white dwarf--brown dwarf binaries and extracting time-resolved observations from them (e.g. \citealt{ben, yifan, rachael}). Initially, the data is sorted into the individual HST orbits so that the direct image for each orbit, which is observed using the F127M filter, is with the relevant spectra which are observed with the G141 grism. Grouping the data in this way before reduction ensures that the correct direct images are used for wavelength calibration, and that the resulting calibration is precise.

Our data was taken using a 256$\times$256 subarray, however \texttt{aXe} is unable to process subarrays properly. To use the \texttt{aXe} pipeline, we first had to pad our data into full-framed arrays which are 1014$\times$1014. To do this we padded the edges of the G141 files and the F127M Direct Images such that the data in our original subarray remains in the centre of the padded array. This allows \texttt{aXe} to use its standard full-frame calibration images during data reduction, and does not alter our actual data.

Any bad pixels in the data are flagged in the data quality (DQ) extension after pre-processing through the \texttt{CalWFC3} pipeline. To correct for these bad pixels, we linearly interpolated neighbouring good pixels to fill in the gaps, using 4 pixels from either side of any bad pixels. We performed this interpolation in both the x-direction and the y-direction, using 16 pixels per interpolation. In addition, we had an extra hot pixel just above the source which was interpolated in a similar way. However, for that hot pixel, we only interpolated in the horizontal direction to avoid accidentally using pixels from the target object in the interpolation. We ignore pixels flagged with cosmic ray hits and use our own cosmic ray detection algorithm which considers the change in count rate between two successive readouts at each pixel to identify cosmic rays. We use a 5$\sigma$ threshold to identify cosmic ray hits which are then interpolated over. After cleaning them, we then performed precise source extraction on our direct images using \texttt{SourceExtractor} \citep{sextractor}.

To prepare the data for extraction using \texttt{aXe}, we executed \texttt{axeprep} on our spectroscopic images. The master sky image \texttt{WFC3.IR.G141.sky.V1.0.fits} from \cite{skyfile} is used to perform optimal background subtraction within the \texttt{axeprep} routine. We found that this background subtraction was not successful in processing our spectra, and produced unphysical results. This is likely due to large contaminant sources in the spectroscopic images of \obj. Thus, we used a manual background subtraction method instead. We created a custom source mask that masked our target source as well as other contaminant sources in the spectroscopic images. After masking all the sources visible in our spectroscopic images, we calculated the median of the resulting background and subtracted that from the entire image. This yielded data with successful background subtraction, which was not affected by contaminant sources.

\begin{figure*}
    \centering
    \includegraphics[width=1.8\columnwidth]{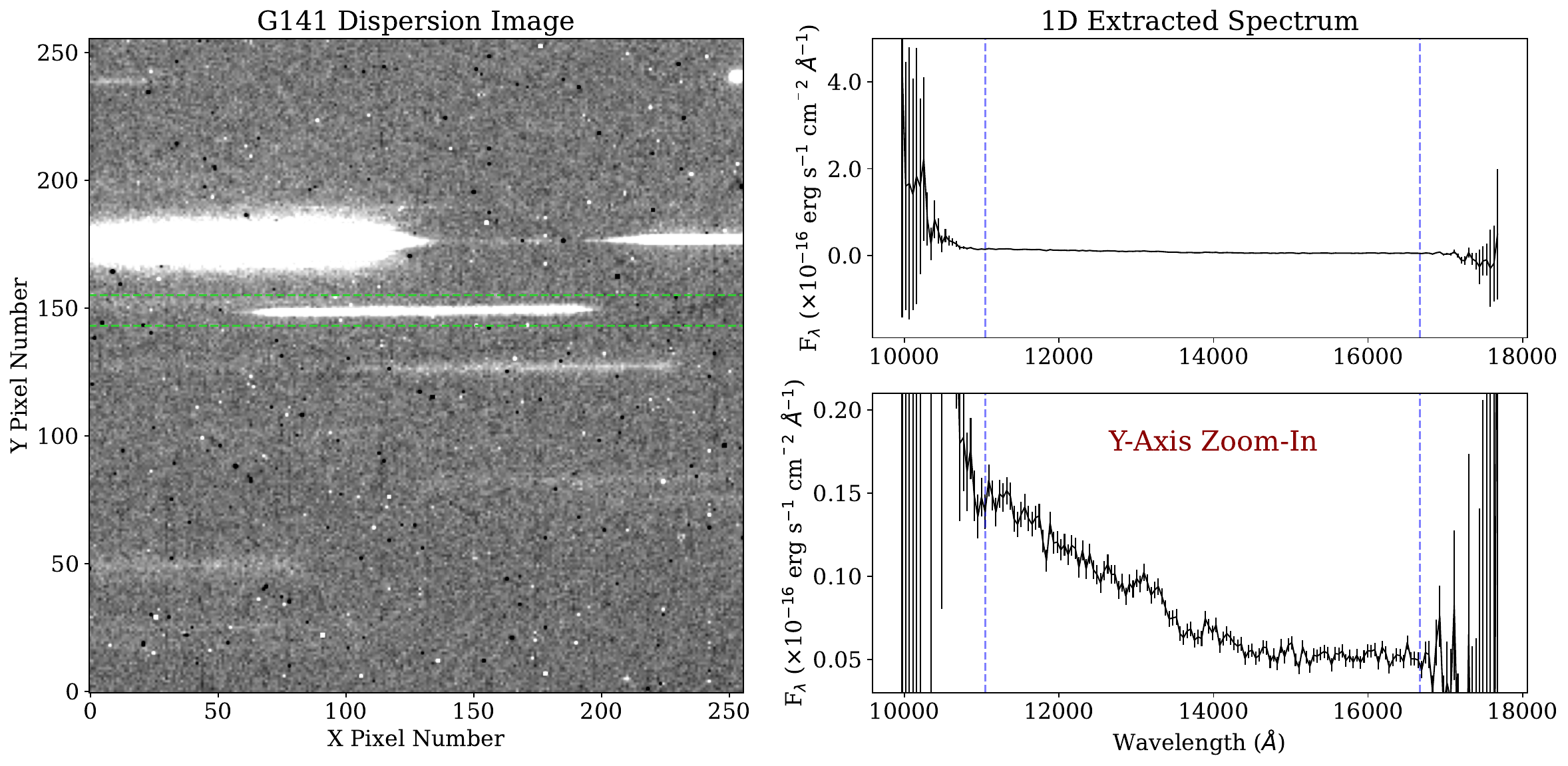}
    \caption{Full reduction performed on one frame of data. The left image is the G141 grism observation, with the grism spectrum of \obj{} highlighted between the green lines. This data is before correcting for cosmic rays and bad pixels. The plots on the right show the data after being passed through \texttt{axeprep} and extracted using \texttt{axecore}. This is the 1D spectrum of a single frame of data. The top right shows the full spectrum, with the blue lines denoting the wavelength range used for the science, spanning 1.1-1.66$\upmu$m. The bottom right graph shows a zoomed in view of this spectrum to highlight the actual shape of our data.}
    \label{fig:extracted}
\end{figure*}

Depending on the detector illumination history and the target fluence rates, data at the beginning of HST orbits can suffer from a ramp-like effect due to charge trapping and a delayed release. To correct for this possibility in our data, we use the RECTE model from \cite{recte} which is a physically motivated charge trap model, however we did not see a visible ramp effect. Figure \ref{fig:extracted} depicts an extracted 1D spectrum from one frame of data in orbit 6. The left hand side shows the G141 spectroscopic image, with the target source highlighted in green, and the right hand side shows the 1D spectrum extracted from that image. This data reduction method was then repeated for each of the 48 individual spectroscopic observations taken by HST. We made wavelength cuts where the flux density errors were over twice the average error between 13000--15000~\AA. Choosing these limits ensures a good signal-to-noise across the spectra, with our clipped data spanning 11000--16600~\AA{} for each spectrum.

\section{Results}
\label{sec:results}

\subsection{Lightcurves}
\label{subsec:lightcurve}
\subsubsection{Creating Lightcurves}

\begin{figure}
    \centering
    \includegraphics[width=\columnwidth]{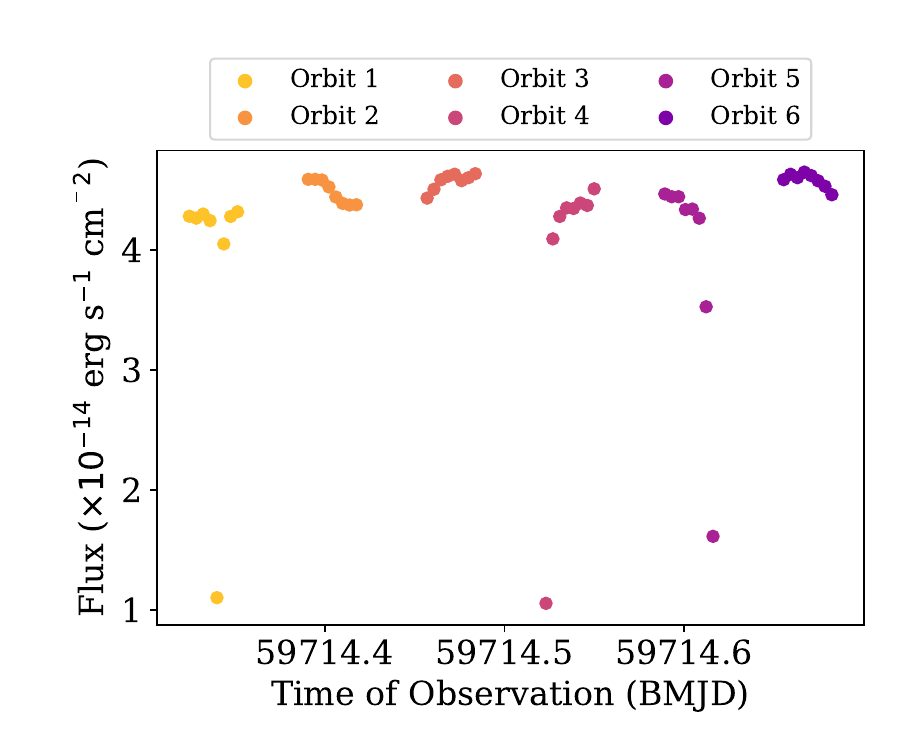}
    \caption{Broadband lightcurve of \obj{} generated by integrating each individual spectroscopic observation between wavelength limits of 11000--16000~\AA. The different colours correspond to each individual orbit of HST that our data spans. The error bars are the same size as the points.}
    \label{fig:lightcurve}
\end{figure}

To derive a lightcurve for \obj{} from our 48 spectra, we integrated the flux density of each spectrum between our wavelength limits. For each spectrum, the integral yields a singular flux point for the lightcurve, and we take the mid-exposure time of that observation as the corresponding time value. We then do this integration for each of the 48 spectra we have, yielding a lightcurve with 48 data points across the 6 consecutive HST orbits. Our lightcurve is shown in Figure \ref{fig:lightcurve}. It captures the eclipse as the brown dwarf fully occults the white dwarf, and we see the non-irradiated nightside of the brown dwarf.

\begin{figure}
    \centering
    \includegraphics[width=\columnwidth]{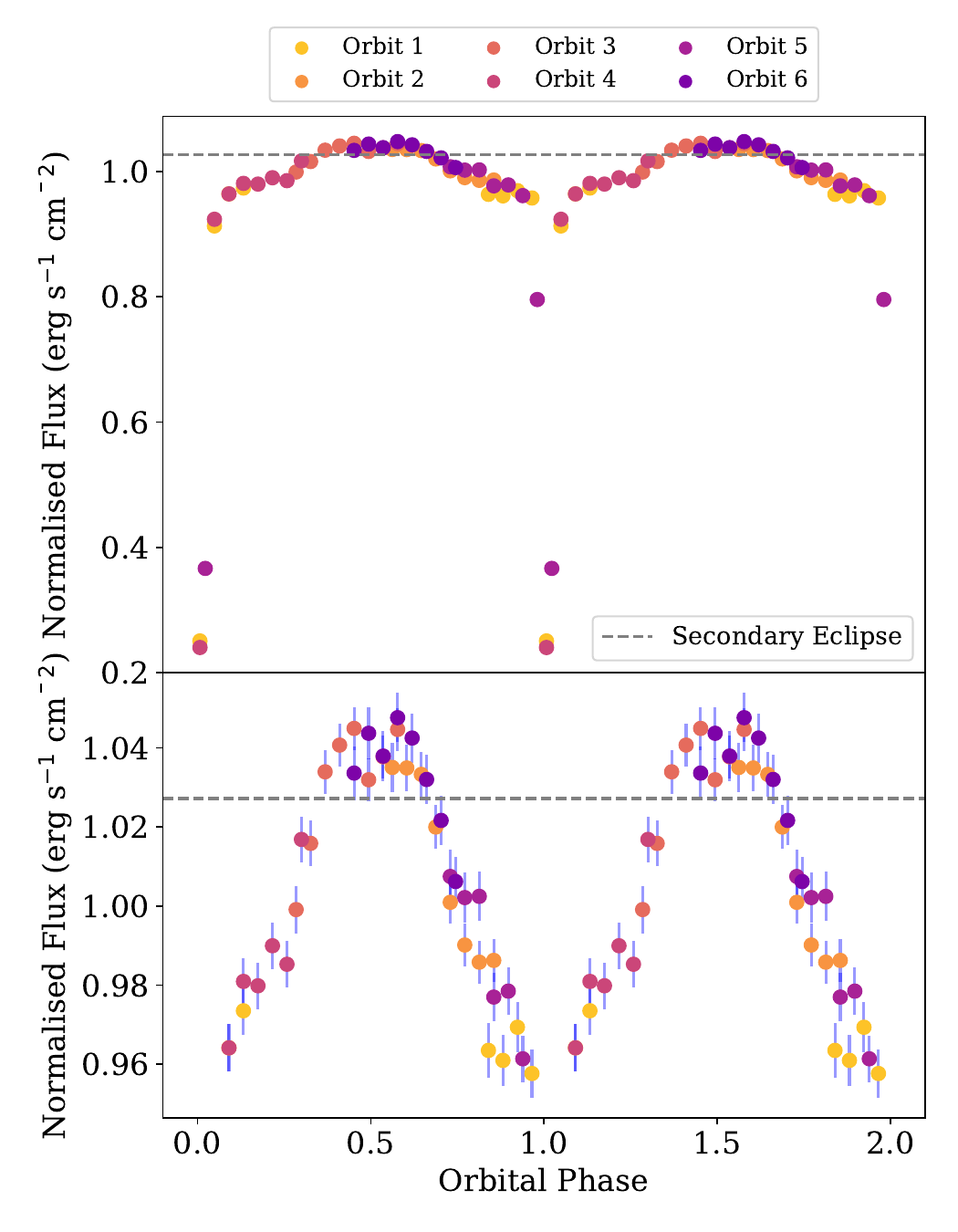}
    \caption{Broadband lightcurve of \obj{} phasefolded on the 0.09155899610~day period to show the variation across the full phase of the orbit. Orbital phase is defined such that a phase of 0 occurs when the brown dwarf eclipses the white dwarf. The data has been repeated to show two orbits of \obj.  The upper panel includes the data points inside the eclipse, ingress and egress. The lower panel has these points removed. The colours correspond to each individual orbit of HST that our data spans. The dashed grey line is the depth at which the secondary eclipse should appear at a phase of 0.5, but this eclipse depth is not observed. The error bars in the upper panel are the same size as the points.}
    \label{fig:phasefold}
\end{figure}

To determine how the lightcurve of \obj{} changes throughout its orbit, we phasefold the lightcurve on its 0.09155899610~day period. Figure \ref{fig:phasefold} shows two complete orbits of \obj, where the data points within the eclipses have been removed in the lower panel. The ephemeris of \obj{} has been included in the phasefolding such that the primary eclipse is at $\phi = 0$ in phase. The dashed grey lines in both panels show the predicted depth of the secondary eclipse based on the radii of the white dwarf and brown dwarf. We do not see a secondary eclipse at $\phi = 0.5$, where the white dwarf transits in front of the brown dwarf. As the predicted secondary eclipse depth is shallow, the inclination of 87.5\degree{} may be sufficient to suppress the eclipse such that it is not detectable outside of the scatter in the lightcurve. As the scatter is large at $\phi = 0.5$, it is not possible to definitively determine the presence of the secondary eclipse. However, since the scatter is larger at this phase than across the rest of the lightcurve, it is likely that the secondary eclipse is present but the inclination is reducing its depth such that it appears as an increase to the scatter rather than a distinct eclipse present below the noise.

There are two data points in our lightcurve that are in the eclipse, which has a duration of 16.6~minutes from the beginning of its ingress to the end of its egress. This is where the brown dwarf fully occults the white dwarf, and the non-irradiated nightside of the brown dwarf is visible. We can therefore expect the spectra of these two observations to be spectra of the nightside of the brown dwarf only, without any contamination flux from the white dwarf. We can then use these solo brown dwarf spectra to remove the white dwarf contribution from the rest of our spectra, which are combinations of the white dwarf and brown dwarf signals (see Section \ref{sec:removingwd}). For the out-of-eclipse lightcurve data, the baseline is not flat and instead shows a sinusoidal shape, which is likely due to the reflection effect. The reflection effect occurs when the dayside atmosphere of the brown dwarf is irradiated by the white dwarf and heats up as a result, absorbing and re-radiating some of the incident flux. This leads to sinusodial brightness variations when viewing the star at different orbital phases \citep{reflection}.

\subsubsection{Lightcurve Fitting}
\label{sec:mcmc}

To verify that the variation we see in the baseline of the lightcurve originates from \objab{} and is not a data systematic, we fit a sinusoidal function to the lightcurve using a Markov Chain Monte Carlo (MCMC) sampler \citep{mcmcgw}. Since we are considering the baseline of the lightcurve and not the eclipse, we remove any point in the eclipse or its ingress and egress.

For our phasefolded lightcurve, we fit the following sinusoidal relationship using MCMC: 
\begin{equation}
   ~~~~~~~~~~~~~~~~~~~~~~y(t) = F_0 + A \sin \Bigg(\frac{2 \pi (t - t_0)}{P} + \Phi_0 \Bigg)~~~,
\label{eq:mcmcmodel}
\end{equation}

where $t_0~=~58381.2439008(10)~$days is the ephemeris and $P~=~0.09155899610(45)$~days is the period, which are both from \cite{wd1032}. Here $t$ is the time of observation, and $F_0$, $A$ and $\Phi_0$ are free parameters to be fit by the MCMC sampler, which correspond to flux offset, amplitude, and phase offset respectively. This method of lightcurve fitting has previously been utilised for fitting the lightcurves of white dwarf--brown dwarf binaries \citep{yifan, rachael}.

For our MCMC models, we imposed uniform priors such that $A > 0$, $-\pi < \Phi_0 \leq \pi$, to cover one full phase with the phase offset. For $F_0$ we used a Gaussian prior with a mean of 1. As the MCMC runs, each walker steps to a new position and calculates the model in Equation \ref{eq:mcmcmodel} using the input parameters at that position. The MCMC sampler then computes the log \textit{likelihood} at that grid position, which is calculated as:

\begin{equation}
    ~~~~~~~~~~\text{ln} (\mathcal{L}) = -\frac{1}{2} \Bigg( \frac{(y_{data} - y_{model})^2}{\sigma_{data}^2} - \text{ln}\Big (\sqrt{2 \pi \sigma_{data}^2} \Big) \Bigg)~~.
\end{equation}

\begin{figure}
    \centering   \includegraphics[width=\columnwidth]{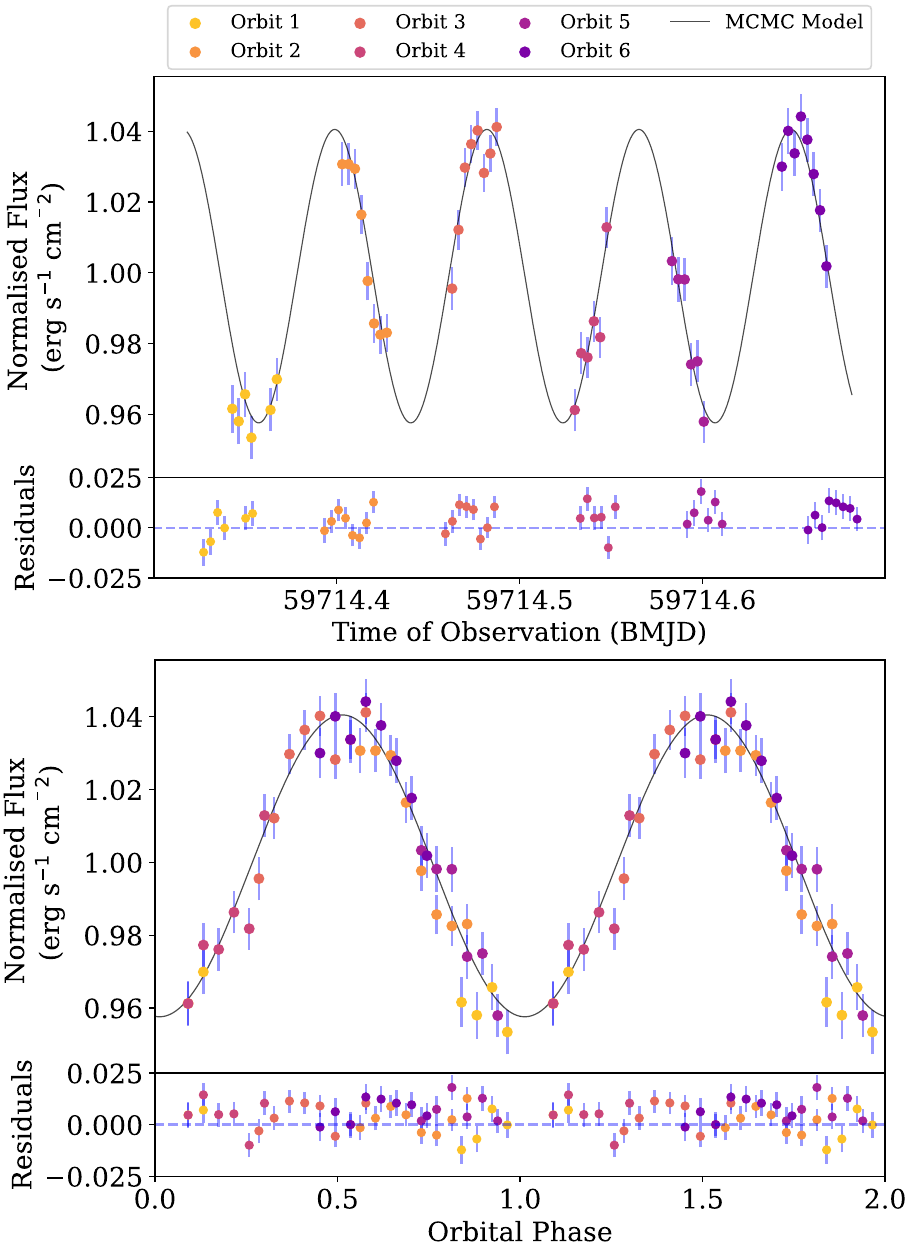}
    \caption{Broadband lightcurve of \obj{} alongside our best-fitting MCMC parameters for the model in Equation \ref{eq:mcmcmodel}. Data points in the eclipse, ingress or egress have been removed. The upper panel shows the full lightcurve with our MCMC model. The lower panel shows the phasefolded lightcurve with the same MCMC model, with the data repeated to show two orbits of the binary. The data has been normalised such that the median is at 1. Our data spans $\sim$4.5 full orbits of \obj.}
    \label{fig:mcmc}
\end{figure}

We use the python \texttt{emcee} package to perform our MCMC fitting \citep{emcee}, with a chain of $N_{steps} = 10,000$, $N_{walkers} = 50$, and a burn-in of 1,000. In Figure \ref{fig:mcmc} we present our lightcurve alongside our best-fitting model from our MCMC analysis, for both our sequential and phasefolded data. The data points in the eclipse, ingress and egress have been removed to properly study the variation in the baseline of the lightcurve.

We find that the first order model in Equation \ref{eq:mcmcmodel} fits our data well. Although we are only able to achieve residuals of $\sim$5\%, we find that adding higher order terms, or adding a cosine term worsens the quality of the fit. This sinusoidal variation in the lightcurve is likely due to the reflection effect, and not other variability. Our best-fit model parameters are $F_0 = (4.442 \pm 0.004) \times 10^{-14}$, $A = (1.84 \pm 0.06) \times 10^{-15}$, and $\Phi_0 = -1.58 \pm 0.03$. Notably, our value for the phase offset, $\Phi_0$ is well within 1$\sigma$ of $\pi /2$. This is expected because we define the eclipse as being at phase $\phi = 0$, and the maximum flux observed is at $\phi = 0.5$, which is where the dayside of the brown dwarf is visible and the white dwarf only blocks a small portion of the brown dwarf flux. As we are fitting a sine function, the maximum would be at an argument of $\pi/2$ but since we define $\phi = 0.5$ as our maximum flux, this would correspond to an argument of $\pi$ in the sine function. Hence, there is a $\pi/2$ offset in the phase, supporting that this variation is due to the reflection effect. This phase offset is consistent with other lightcurves of irradiated brown dwarfs obtained via HST observations (e.g. \citealt{ben, yifan}).

\subsubsection{Sub-band Lightcurves}

\begin{figure}
    \centering
    \includegraphics[width=\columnwidth]{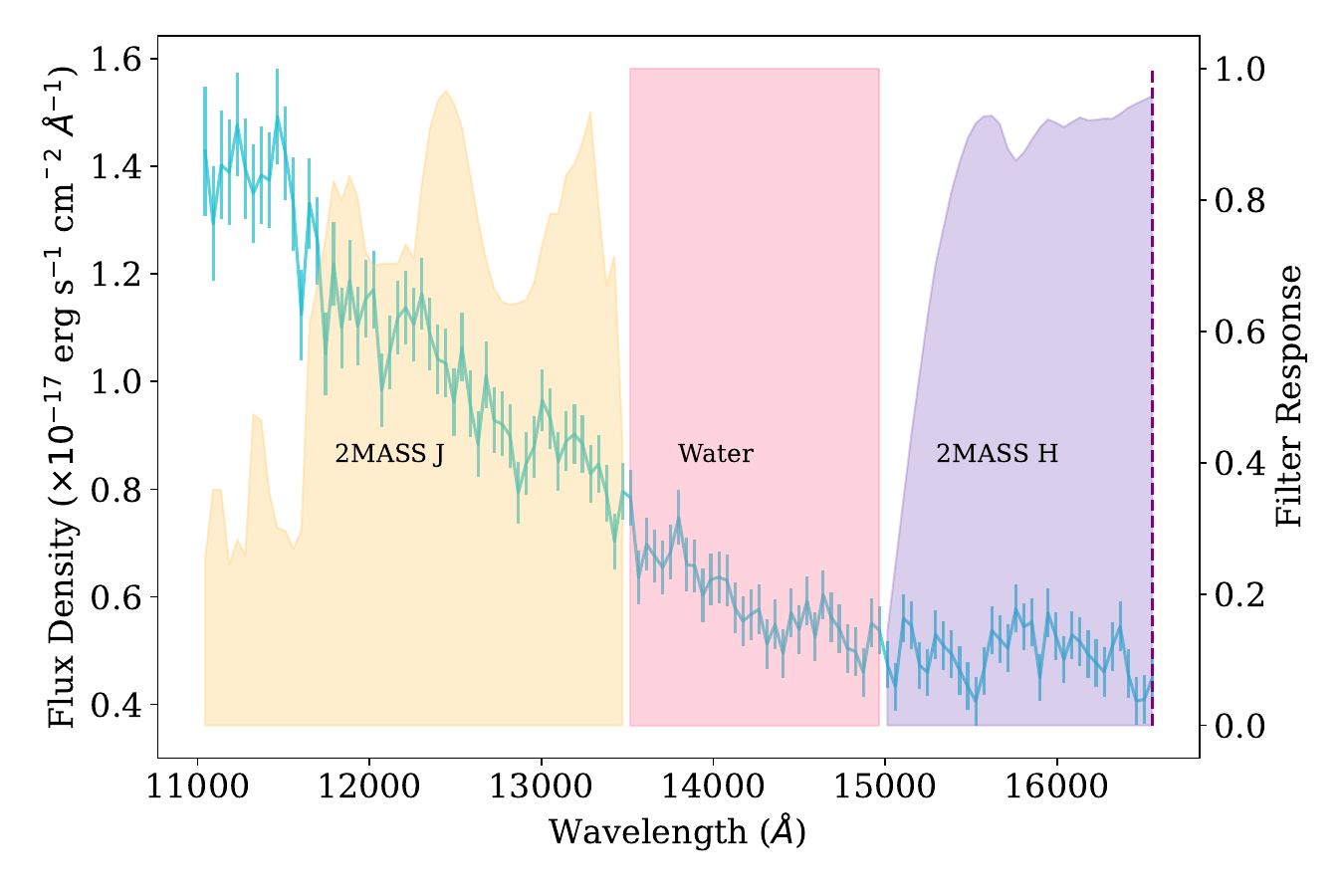}
    \caption{Spectrum of \obj{} after data reduction. The filter transmission profiles for 2MASS $J$, water, and 2MASS $K$ filters used to create the sub-band lightcurves are shown in yellow, pink and purple respectively. The filter profiles were resampled to match the resolution of our data before multiplying them with our spectrum.}
    \label{fig:filters}
\end{figure}

The varying dust and gas opacities within brown dwarf and exoplanet atmospheres cause different wavelengths to probe different pressure regions within the atmosphere \citep{buenzli, ben}. Analysing lightcurves in individual $J$-, water and $H$-wavebands has previously identified wavelength-dependent intensity changes in brown dwarfs, indicating the presence of atmospheric structure and dynamics which are pressure-dependent (e.g. \citealt{apai, rachael}). To identify any wavelength-dependent changes in intensity in our lightcurve, we generate sub-band lightcurves from our observations.  By comparing any changes in flux across different wavelength regions, we can quantify any differences that would indicate pressure-dependent behaviour in the atmosphere. We derive three separate sub-band lightcurves for different sections of our wavelength range, spanning a $J$ filter, a water filter, and a modified $H$ filter. These filter choices have previously recovered wavelength-dependent intensity variations in the atmospheres of brown dwarfs. Figure \ref{fig:filters} depicts our overall combined spectrum for \obj{} with the filters we use to generate these sub-band lightcurves overlaid.

\begin{figure}
    \centering
    \includegraphics[width=\columnwidth]{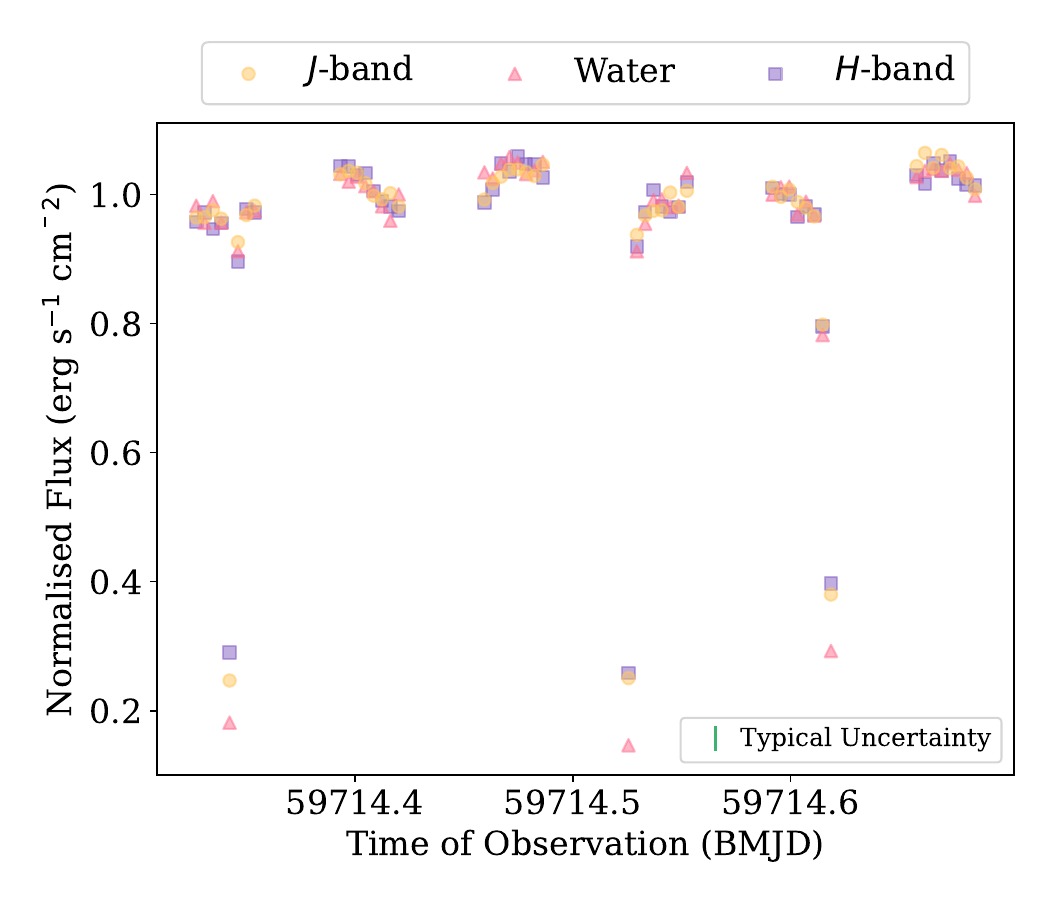}
    \caption{Normalised sub-band lightcurves of \obj{} generated using the 2MASS $J$-, water and 2MASS $H$-bands by integrating spectra between 11000--13500~\AA, 13500--14500~\AA{} and 14500--16600~\AA ~respectively. The yellow circles, pink triangles and purple squares correspond to the 2MASS $J$-, water, and 2MASS $H$-bands respectively. The typical uncertainty size is shown in the lower right corner.}
    \label{fig:splitlc}
\end{figure}

To create a lightcurve in the $J$-band, we used the filter transmission from the Two-Micron All Sky Survey (2MASS) $J$ filter \citep{2massfilter}, and multiplied this with our spectra between 11000~\AA~and 13500~\AA. We then integrated the resulting spectrum between these wavelengths to obtain a single flux point. Similarly for the $H$-band, we multiplied our spectrum with the filter transmission for the 2MASS $H$ filter beginning at 14500~\AA, and then integrated. Since the 2MASS $H$ filter extends beyond our quality data at 16600~\AA, we modified this $H$ filter to end at the same point as our data. Water absorption is an important feature in the atmospheres of brown dwarfs, so we created a top hat-shaped water filter spanning 13500--14500~\AA. We then integrated between these wavelengths to create a water-band lightcurve. After applying this methodology to each of our 48 individual spectra, we normalised the median of each lightcurve to 1 to obtain the sub-band lightcurves depicted in Figure \ref{fig:splitlc}.

Unlike those seen in NLTT5306B \citep{rachael}, we do not find any significant differences between the sub-band lightcurves in different bands, indicating that there is no evidence of a pressure-dependent dayside-nightside temperature contrast or dynamics in the atmosphere of \objb{}. The only difference present is that the eclipses in our lightcurve are deeper in the water band. This is not unexpected as the thermal structures on the nightside of the brown dwarf are such that temperature decreases with decreasing pressure. The water band then probes the cooler upper layers in the nightside of the brown dwarf. The apparent strength of the water absorption features in the system are also greater in the eclipse when the brown dwarf occults the white dwarf.

\begin{figure}
    \centering
    \includegraphics[width=\columnwidth]{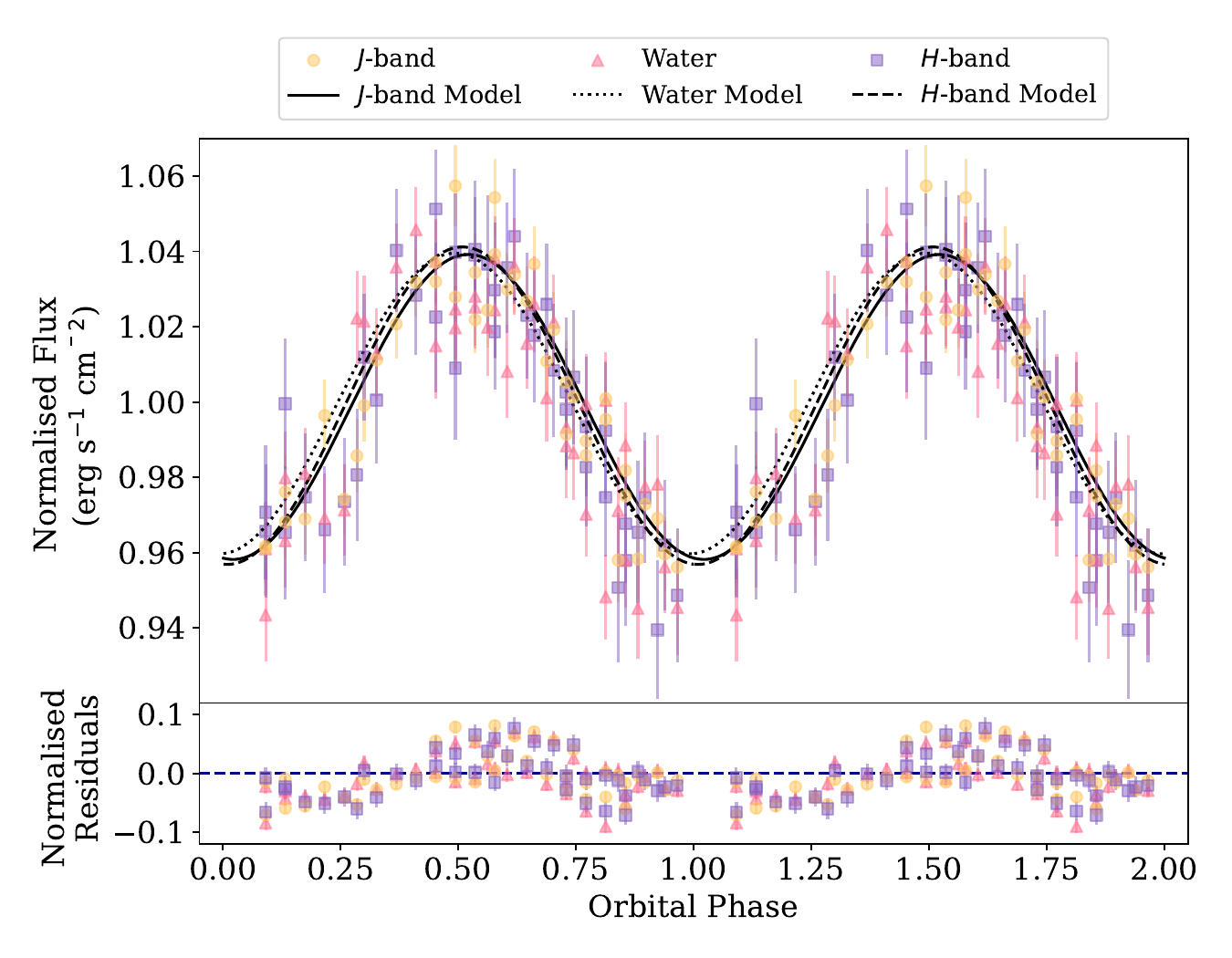}
    \caption{Phasefolded lightcurves in the $J$-, water and $H$-bands alongside the best-fit models from our MCMC analysis, following \protect\cite{rachael}. The data has been normalised such that the median is at 1. The $J$-band is depicted with yellow circles and a solid line, the water band has pink triangles with a dotted line, and the $H$-band is shown with purple squares and a dashed line. The data has been repeated to show two orbits of \obj{}.}
    \label{fig:splitmcmc}
\end{figure}

To fit the baseline of our $J$-, water and $H$-band lightcurves, we followed the same procedure described in Section \ref{sec:mcmc} for each individual lightcurve. We find that all of the $J$-, water and $H$-band lightcurves are well-fit with the first order MCMC model we use. The MCMC fitting for our phasefolded sub-band lightcurves is shown in Figure \ref{fig:splitmcmc}. We note that the flux uncertainties are larger for our sub-band lightcurves compared to our broadband lightcurve, which is due to the smaller samples of data used to calculated each lightcurve point. We do not see any significant differences in the peak-to-peak amplitude or phase offset between the MCMC fitting for each wavelength band. With the eclipse points removed, the peak-to-peak amplitude of each sub-band lightcurve are within 0.4$\sigma$ of each other, where $\sigma$ is the average flux uncertainty for each band. Additionally, there is only a minimal phase offset present between the best-fitting models for the $J$-, water and $H$-band lightcurves, and these offsets are within the resolution of the intervals in orbital phase, which is $\sim$0.04. There are no significant amplitude variations or phase offsets between the wavebands. The consistent phase offsets between the bands indicates the absence of jets which enable efficient heat redistribution from the dayside to the nightside. As such, the poor heat redistribution would lead to an observed temperature contrast between the two hemispheres. This also indicates that the observed reflection effect is not dependent on the pressure in the atmosphere. It is therefore likely that the irradiation the brown dwarf receives from the white dwarf is equally penetrating the entire atmosphere, as opposed to only affecting certain depths. 

There is an apparent scatter in the eclipse depths at different observation times seen in Figure \ref{fig:splitlc}. Considering the two lightcurve points in the centre of the eclipse, where the nightside of the brown dwarf is observed, rather than its ingress or egress, this variation in the eclipse depth is only marginally higher than the average scatter of the lightcurves. The maximum orbit-to-orbit eclipse depth variation is seen in the water band, with a difference of 1.19$\sigma$, where $\sigma$ is the average scatter between the data and the best-fitting MCMC model of the reflection effect. As this remaining variation is so small, it is unlikely that there is any inherent variability present within \objb{}.

\subsection{Spectra}
\label{subsec:spectra}

\begin{figure}
    \centering
    \includegraphics[width=\columnwidth]{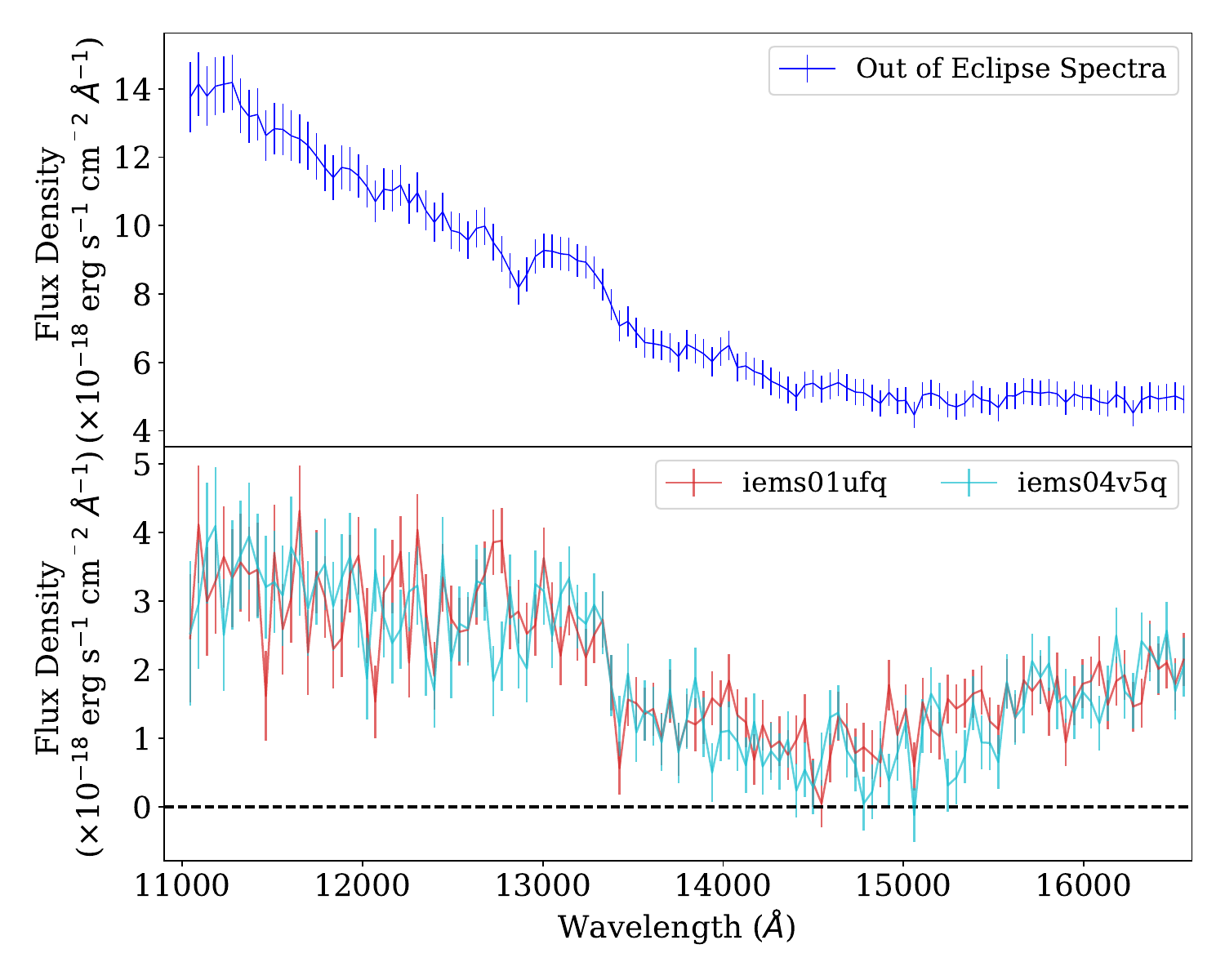}
    \caption{WFC3 spectra of \obj{} showing spectra in and outside the eclipse. The upper panel shows the combined out of eclipse spectra, which is a combination of the contributions from both the white dwarf and the brown dwarf. The lower panel shows the two spectra within the eclipse, which are the nightside of the brown dwarf alone.}
    \label{fig:eclipsewithorbit}
\end{figure}

We combined our 48 individual spectra first for each orbit, and then we combined all orbits together. Figure \ref{fig:eclipsewithorbit} shows our average spectrum across all 6 orbits in the upper panel. During this process, we identified two frames that were significantly different to the rest, which are shown in the lower panel of Figure \ref{fig:eclipsewithorbit}. These different spectra are the two data points within the eclipse in our broadband lightcurves, when we are only observing the nightside of the brown dwarf, hence their lower flux density and different spectral features. These two spectra were excluded from the combining, and instead we created a combined in-eclipse spectrum using them, so that we have an in-eclipse spectrum, and an out-of-eclipse spectrum. The in-eclipse spectra were obtained entirely in the eclipse and do not contain any data taken in the ingress or egress. The in-eclipse spectra span phases of 0.961--1.0 and 0.0--0.0006, with the primary eclipse of the binary spanning phases of 0.9--1.0 and 0.0--0.1 from the K2 lightcurve presented in \cite{wd1032}.

\subsubsection{Removing the White Dwarf Contribution}
\label{sec:removingwd}

Since our observations are combined spectra of the white dwarf and the brown dwarf, we need to isolate the brown dwarf signal in order to study its phase-resolved spectra. As \objab{} is eclipsing and we have identified 2 spectra that are in the eclipse, we already have spectra of the nightside of the brown dwarf alone. If we consider the night phase of \objab{} where the brown dwarf eclipses the white dwarf, at orbital phases surrounding $\phi = 0$ such that the white dwarf is not fully occulted, the flux received from the system is

\begin{equation}
    ~~~~~~~~~~~~~~~~~~~~~~~~~~~~~~~~~~~F_{\text{night}} = F_{\text{WD}} + F_{\text{nightBD}}~~,
\label{eq:nightphase}
\end{equation}

where $F_{\text{night}}$ is the total flux observed from \objab{} during the night phase, $F_{\text{WD}}$ is the flux emitted from the white dwarf, and $F_{\text{nightBD}}$ is the flux emitted by the nightside of the brown dwarf. Since the eclipse spectrum of the brown dwarf is equivalent to its nightside emission, we can calculate the flux from the white dwarf only as

\begin{equation}
    ~~~~~~~~~~~~~~~~~~~~~~~~~~~~~~~~~~~F_{\text{WD}} = F_{\text{night}} - F_{\text{eclipseBD}}~~.
    \label{eq:wdspec}
\end{equation}

\begin{figure}
    \centering
    \includegraphics[width=\columnwidth]{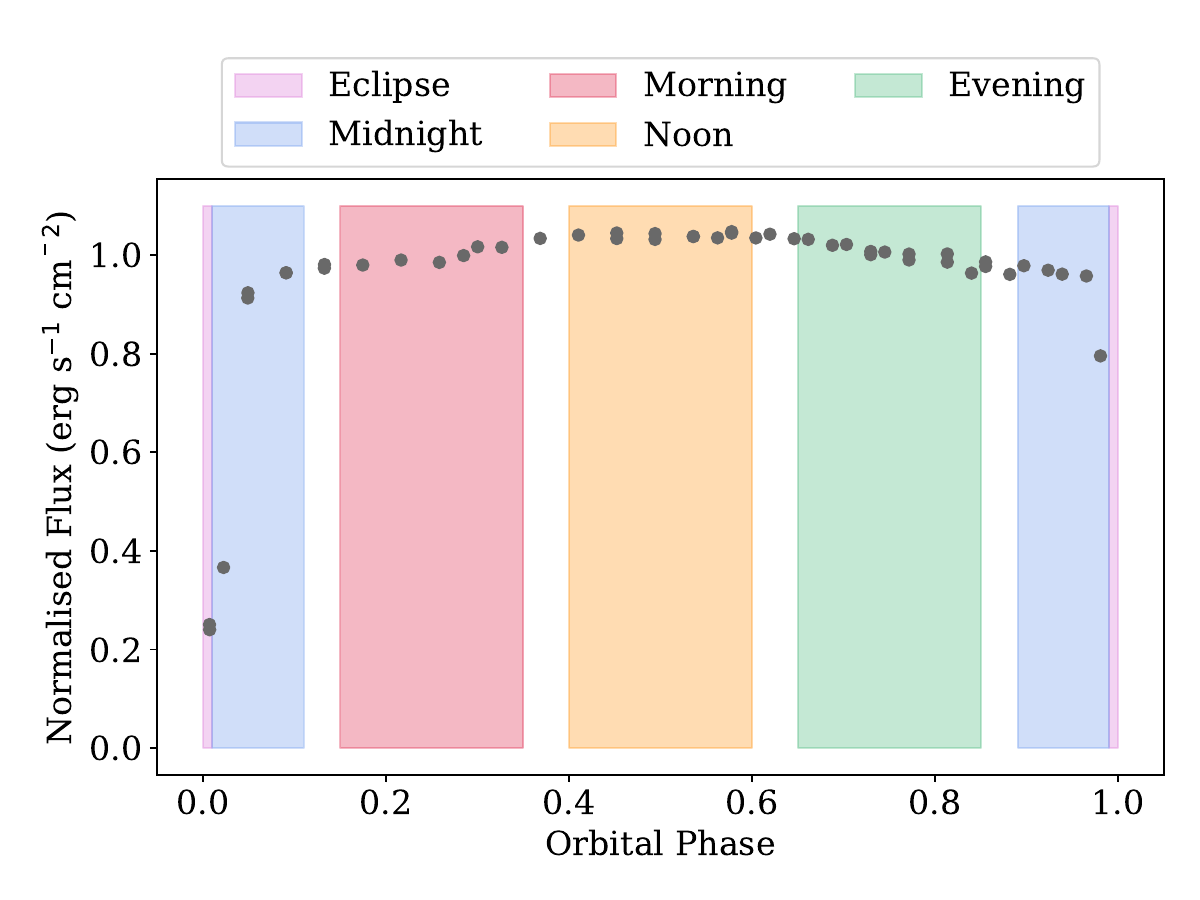}
    \caption{Phasefolded lightcurve of \obj{} showing a single orbit with phase windows overplotted. The phase windows are defined as eclipse: $\phi = 0.00-0.01$, and $\phi = 0.99-1.00$, midnight:  $\phi = 0.89-0.99$ and $\phi = 0.01-0.11$, morning: $\phi = 0.15-0.35$, noon: $\phi = 0.40-0.60$, evening: $\phi = 0.65-0.85$. These windows are shown in pink, blue, red, yellow and green respectively.}
    \label{fig:phasewindows}
\end{figure}

To extract the spectrum of the white dwarf from our combined white dwarf--brown dwarf spectra, we defined a series of phase windows for the four quarters of our full orbital phase. The phase windows are defined as eclipse: $\phi = 0.00-0.01$, and $\phi = 0.99-1.00$, midnight, that is the rest of the night phase which includes the ingress and egress but does not include the eclipse: $\phi = 0.89-0.99$ and $\phi = 0.01-0.11$, noon, when the white dwarf eclipses the brown dwarf: $\phi = 0.40-0.60$, morning: $\phi = 0.15-0.35$, and evening: $\phi = 0.65-0.85$. In the morning and evening phases, the white dwarf and the brown dwarf are both side on with respect to the observer. Figure \ref{fig:phasewindows} shows the normalised, phasefolded lightcurve for \obj{} alongside these phase windows.

For each phase window, we combined the spectra within that window to create single averaged spectra for each of eclipse, midnight, morning, noon, and evening. We then subtracted our combined eclipse spectrum from our midnight spectrum following Equation \ref{eq:wdspec} to obtain our spectrum of the white dwarf alone.

We also modelled the spectrum of the white dwarf in \obj{} using the \cite{koester} DA white dwarf models. From this model grid which uses $T_{\text{eff}}$ and log~$g$ as free parameters, we bi-linearly interpolate along both axes in the grid to create a model for our parameters of \obja{}, $T_{\text{eff}} = 9950 \pm 150~$K and log~$g = 7.65 \pm 0.13$.

To account for the uncertainties in our values of $T_{\text{eff}}$ and log~$g$, we generated flux uncertainties for our bi-linearly interpolated Koester model using a series of Gaussians. We first created Gaussian distributions centred on the measured values of $T_{\text{eff}} = 9950$~K and log~$g = 7.65$ using their uncertainties, 150~K and 0.13, as the respective standard deviations. We sampled 10,000 unique parameter pairs of ($T_{\text{eff}}$, log~$g$) from these distributions and then generated 10,000 white dwarf models via bi-linear interpolation of the Koester model grid.

\begin{figure}
    \centering
    \includegraphics[width=0.9\columnwidth]{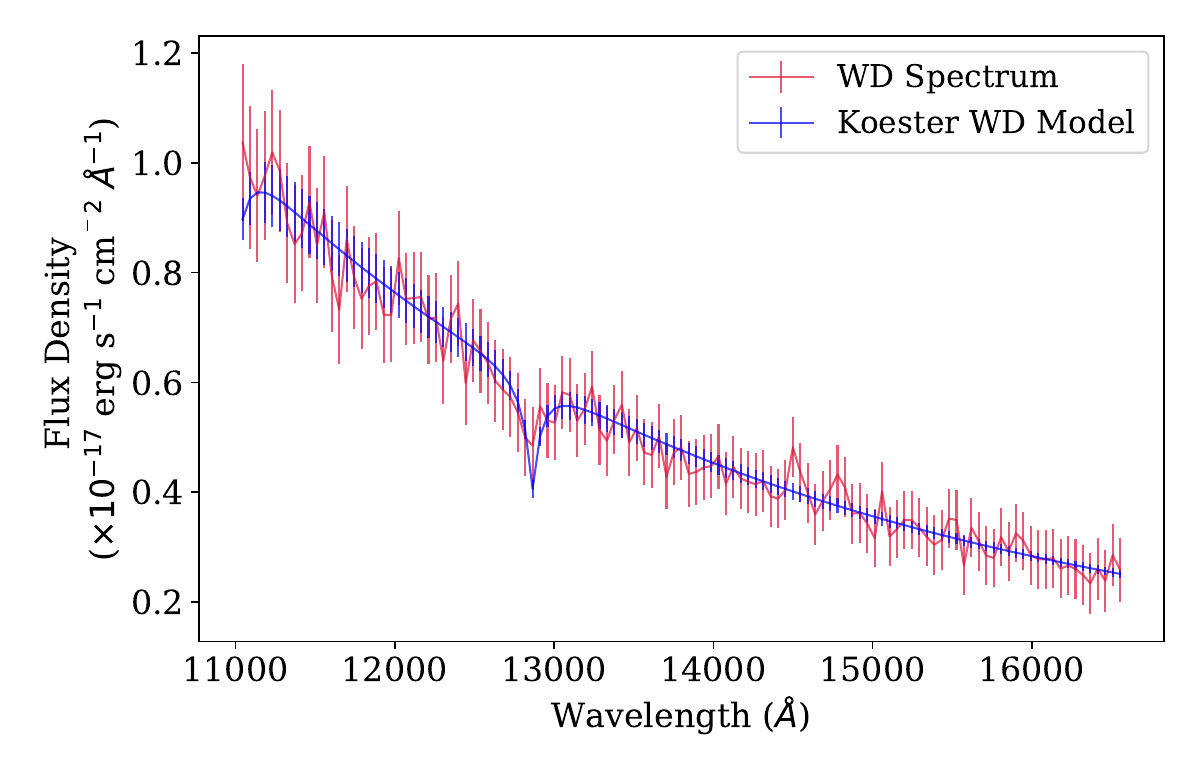}
    \caption{Comparison of our extracted white dwarf spectrum in red with our white dwarf Koester model in blue, with flux uncertainties determined using our Gaussian methodology. The Koester model has been scaled to our derived white dwarf spectrum.}
    \label{fig:koesterwithspec}
\end{figure}

To determine the flux uncertainties of our white dwarf model, we calculated the Full Width Half Maximum of the Gaussian formed by the fluxes from our 10,000 models at each wavelength point. To scale our white dwarf model to the white dwarf spectrum we have extracted, we multiply by the scale factor $\Big( \frac{R_{\text{WD}}}{D} \Big)^2$, where the white dwarf radius, $R_{\text{WD}}$, and distance, $D$, are those from \cite{wd1032}. We find a constant 10\% flux offset between our white dwarf spectrum and our scaled Koester model, but this is well within the 24\% uncertainty in the scale factor, which is dominated by the uncertainty in the $Gaia$ DR3 distance to \obj. Figure \ref{fig:koesterwithspec} compares the white dwarf spectrum we extracted from our observations with our Koester model that has been scaled by the scale factor, and additionally scaled to correct for the 10\% flux offset. Our rescaled Koester model and white dwarf spectrum are very well-aligned with each other, with the initial flux offset well within the uncertainty in our scale factor, indicating that we have successfully extracted the white dwarf spectrum, and our data is consistent with the white dwarf radius and distance presented in \cite{wd1032}.

\subsubsection{Phase Resolved Brown Dwarf Spectra}
\label{sec:phasespec}

\begin{figure}
    \centering    
    \includegraphics[width=\columnwidth]{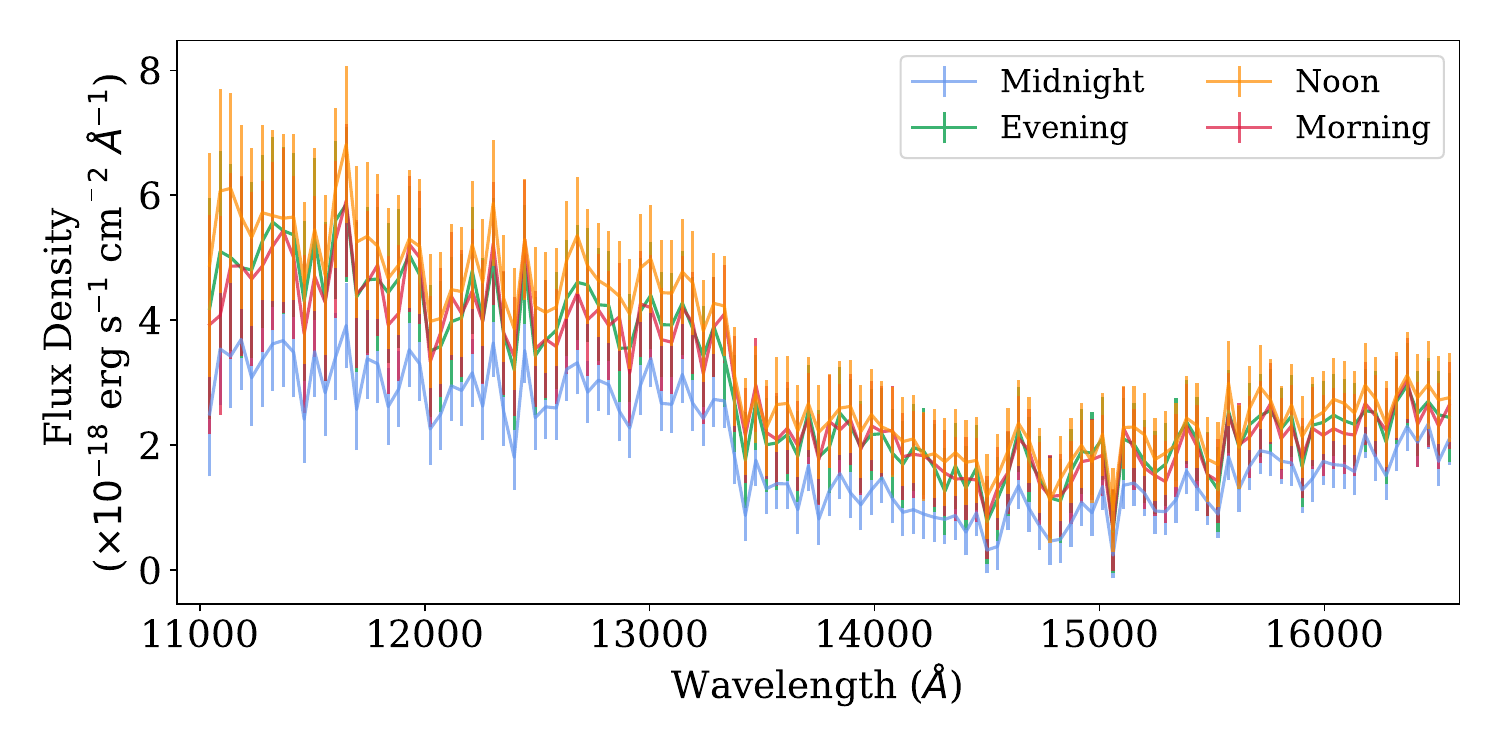}
    \caption{Phase-resolved spectra of the brown dwarf in \obj. The phases considered are the midnight phase (blue), morning (red), noon (orange) and evening (green). Here the midnight spectrum is the nightside of the brown dwarf, and the noon spectrum is its dayside.}
    \label{fig:phasespec}
\end{figure}

We subtracted our isolated spectrum of the white dwarf from the combined white dwarf + brown dwarf spectra we had in each phase window of midnight, morning, noon, and evening. Our phase-resolved spectra of the brown dwarf are depicted in Figure \ref{fig:phasespec}, with the midnight spectrum hereafter corresponding to the eclipse spectrum of the nightside of the brown dwarf, observed at $\phi = 0$. The large feature in all of the spectra at $\sim$13500~\AA{} is due to water absorption in the brown dwarf atmosphere. The eclipse (midnight) spectrum is the faintest, which is expected due to only the nightside of the brown dwarf being visible. Similarly, the noon spectrum is the brightest as this is when the irradiated dayside of the brown dwarf is visible. We have corrected our noon spectrum to account for the small fraction of flux that is blocked as the white dwarf transits the brown dwarf. The morning and evening spectra are in-between these two extremes and are incredibly similar to each other. This is expected as at morning and evening the brown dwarf appears half-irradiated and half-non-irradiated to the observer. We find that on average, our dayside (noon) spectrum is 81\% brighter than our nightside (midnight) spectrum. This indicates a high level of irradiation from the white dwarf primary coupled with poor heat redistribution between the irradiated and non-irradiated hemispheres of the brown dwarf \citep{heatredist}.

\subsection{Brightness Temperature}
\label{subsec:brighttemp}

The thermal structure of the atmosphere of \objb{} will be influenced by the differing opacities within the atmosphere, the internal heat flux, and the absorbed flux due to irradiation from the white dwarf \citep{thermstruc}. To identify potential differences between the thermal structure of the dayside and nightside hemispheres of \objb{}, we investigated its brightness temperature. The brightness temperature is the temperature a blackbody of the same radius would be at to emit the flux that we observe.

\begin{figure}
    \centering    
    \includegraphics[width=\columnwidth]{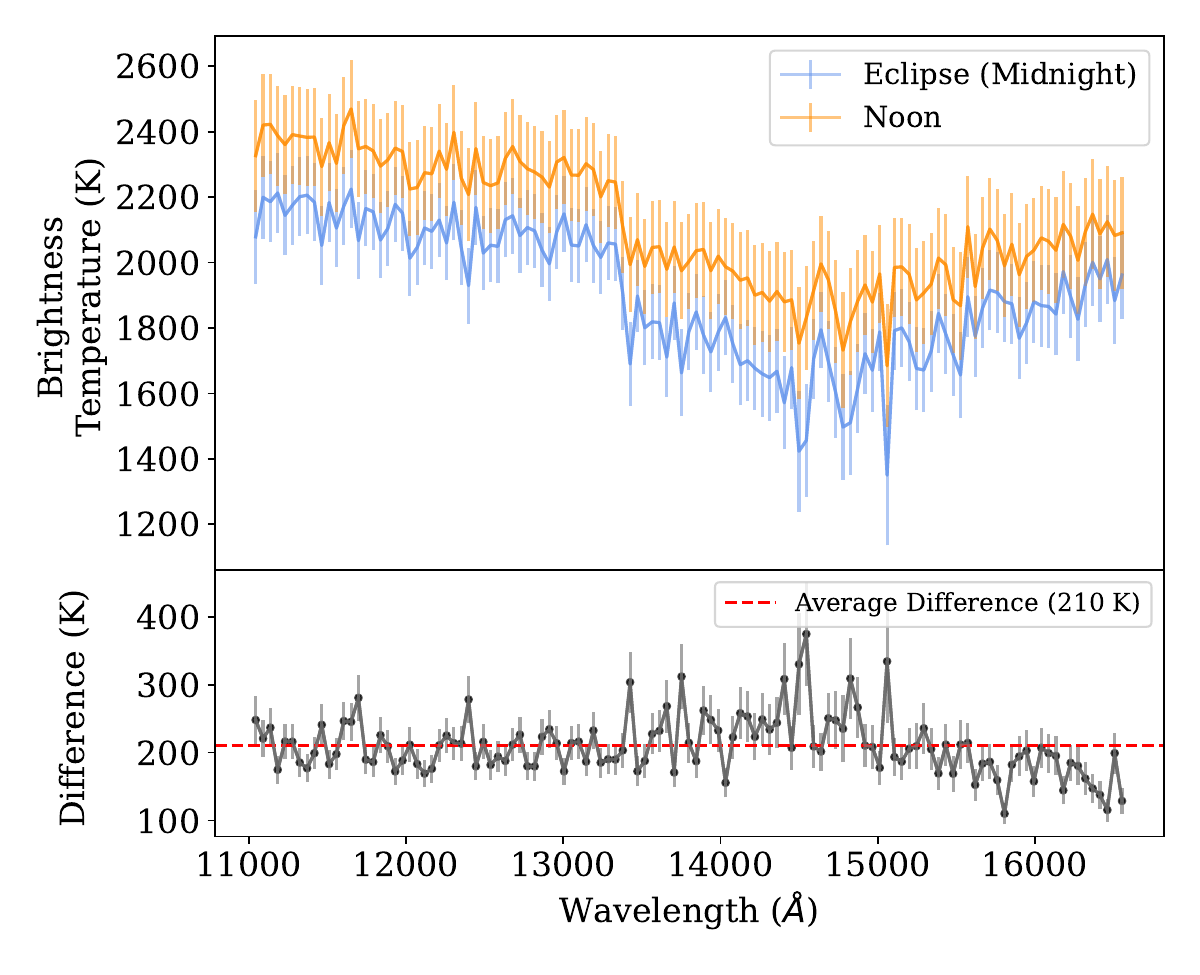}
    \caption{Brightness temperature of \objb{} for both the dayside and nightside, determined by calculating the blackbody temperature from the Planck equation for each wavelength. The difference between the nightside and the dayside is near-constant across the entire spectral range at 210~K}
    \label{fig:brighttemp}
\end{figure}

We calculate the brightness temperature using Planck's law of radiation for a blackbody with our observed flux, the distance of 313~pc to \obj{}, and the radius of the brown dwarf, which is 1.024~R$_{\text{Jup}}$. We calculate the brightness temperature at each wavelength point for both our dayside and nightside spectra, which is shown in Figure \ref{fig:brighttemp}.

The brightness temperatures range between $\sim$1400~K and 2500~K, and show a strong wavelength dependence in both the dayside and nightside hemispheres. The lowest brightness temperatures for both the dayside and the nightside are at $\sim$13500--14500~\AA, which is the water absorption band. Our brightness temperatures show a temperature difference of 210~K between the dayside and nightside of the brown dwarf across the full spectral range. Comparing the brightness temperatures on the dayside and nightside, the vast majority have a difference within 2$\sigma$ of the overall 210~K temperature contrast, with only a few deviations concentrated around the water absorption feature. As our temperature difference is nearly constant, it is likely that the atmosphere in both hemispheres has the same composition and opacity sources. However, there are small changes in the shape of the brightness temperature between the different hemispheres, such as the nightside exhibiting deeper features around 14500--15000~\AA{}, which could result from the dissociation of atmospheric molecules in the dayside of the brown dwarf due to the irradiation it receives, and indicate that the response time is of a similar order to the orbital period.

\section{Comparison to Field Brown Dwarfs}
\label{sec:fieldbds}

\begin{figure}
    \centering    
    \includegraphics[width=\columnwidth]{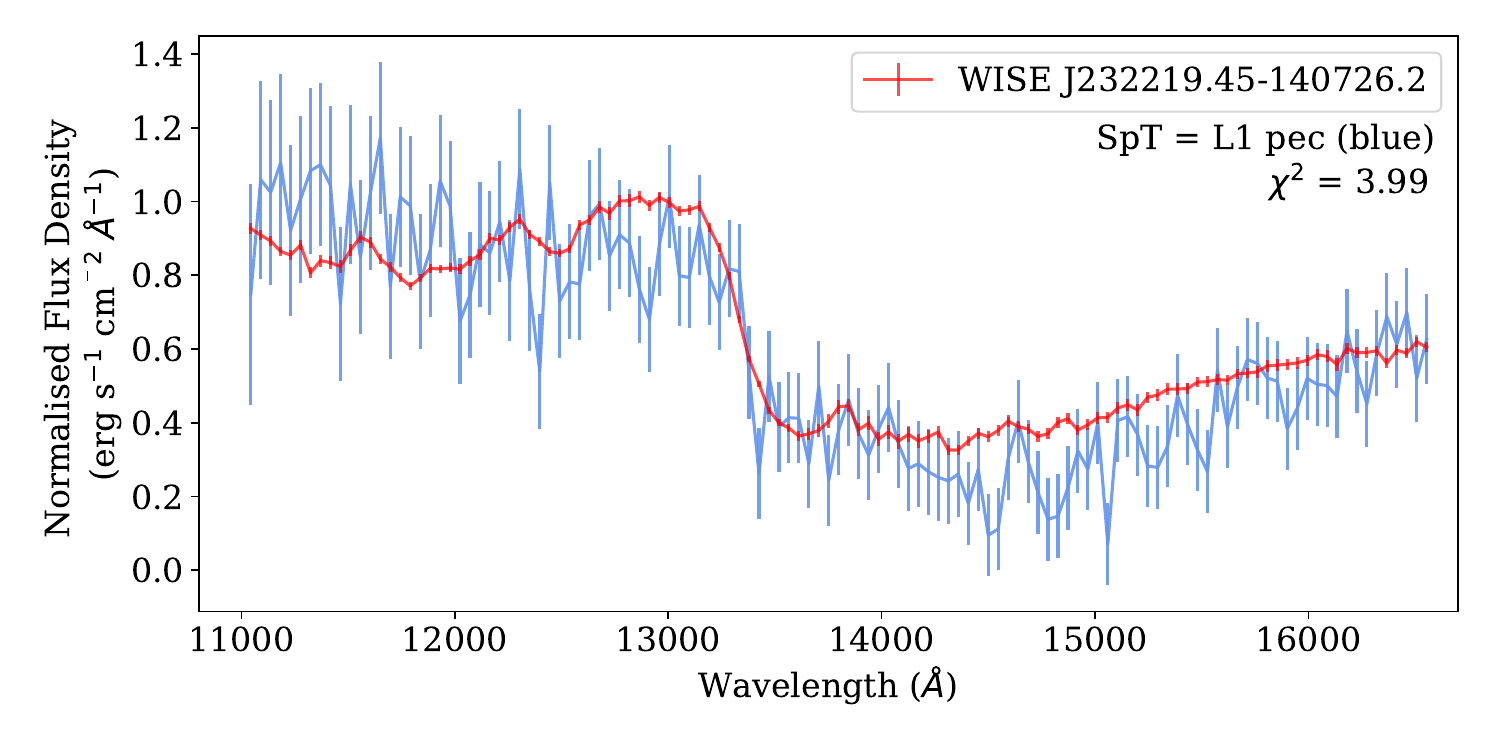}
    \caption{Comparison of the nightside spectrum of the brown dwarf with field brown dwarfs using $\chi^2$ fitting. The red line shows the best-fit object for the SpeX library, WISE~J323319.45-140726.2 which is an L1 pec spectral type with a $\chi^2$ of 3.99. Our nightside spectrum is shown in blue.}
    \label{fig:chisq}
\end{figure}

To identify the spectral type of the brown dwarf companion in \obj, we compare our nightside spectrum to spectral libraries of non-irradiated field brown dwarfs. We used two different databases for this, the SpeX Prism database \citep{spexlib}, and the Cloud Atlas spectral library \citep{cloudatlas}. The SpeX prism library comprises 234 L, T and Y dwarf spectra with available data, all observed using SpeX, which is a ground-based spectrograph. The Cloud Atlas library contains 53 usable L, T and Y dwarf spectra that have been observed with the HST WFC3 instrument. To scale our spectrum of \objb{} to the same flux as the field brown dwarfs, we normalise the flux at 13000~\AA{} to 1. This wavelength probes high pressure areas in the atmosphere and should thus be less affected by irradiation \citep{rachael}. To determine which field brown dwarf spectrum best matches our nightside spectrum, we resample the field spectra to the same resolution as our data, and calculate the $\chi^2$ value between our data and each field brown dwarf spectrum. Figure \ref{fig:chisq} shows the best fitting field brown dwarf from the SpeX and Cloud Atlas sources.

We note that the $\chi^2$ values were consistently lower for the SpeX template spectra than for the Cloud Atlas ones. Additionally, although the Cloud Atlas library is HST WFC3 data, it only starts at a spectral type of L4.5, whereas the SpeX library starts at L0. \objb{} is outside of the spectral type coverage of the Cloud Atlas Library, but it is included in the coverage of the SpeX database. Our best-fitting spectrum is WISE~J323319.45-140726.2 from the SpeX prism library, which is an L1 pec spectral type brown dwarf. L1 peculiar brown dwarfs tend to have deeper water absorption and a bluer slope longwards of 13000~\AA{} compared to their L1 counterparts \citep{L1pec}. A spectral type of L1 pec is also consistent with our average nightside brightness temperature of 1909~K. 

\section{Atmospheric Models}
\label{sec:atmosmodels}

\subsection{Non-Irradiated Brown Dwarf Models}
\label{subsec:nonirrbdmodels}

We compare our nightside and irradiated dayside spectra of \objb{} to the ATMO 2020 suite of non-irradiated atmosphere models, which do not include cloud opacity, designed for brown dwarfs and giant exoplanets from \cite{atmo2020}. The models have a solar metallicity and vary through log $g$ = 2.5--5.5 in steps of 0.5, and $T_{\text{eff}}$ = 200--3000~K, with steps of 50~K and 100~K Kelvin lower and higher than $T_{\text{eff}}$ = 600~K respectively. The models are generated by the \texttt{ATMO} code which solves the pressure-temperature structure of an atmosphere following a radiative-convective equilibrium model. Since these models do not consider irradiation, the effective temperature is equivalent to the internal heat flux. There are three model grids, one for equlibrium chemistry, and two for disequilibrium chemistry with different strengths of vertical mixing, which is characterised by the $K_{zz}$ parameter.

\begin{figure}
    \centering    
    \includegraphics[width=\columnwidth]{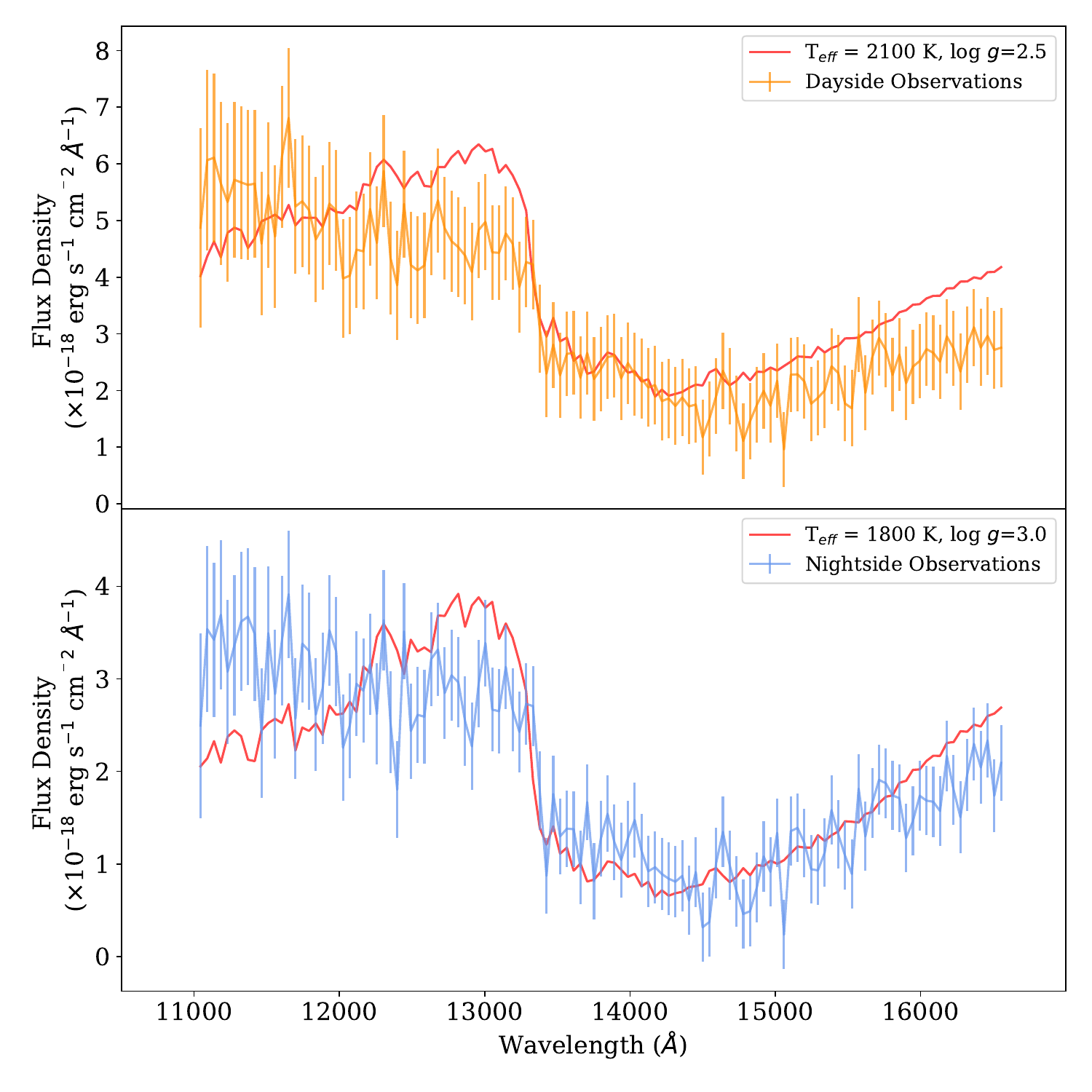}
    \caption{Comparison of our spectra of \objb{} with chemical equilibrium ATMO 2020 models using $\chi^2$ fitting. The upper panel shows our dayside spectrum in orange, with the red line depicting the best-fit ATMO model which has $T_{\text{eff}} = 2100$~K and log $g$ = 2.5. The lower panel shows our nightside spectrum in light blue, with the red line depicting the best-fit ATMO model which has $T_{\text{eff}} = 1800$~K and log $g$ = 3.0.}
    \label{fig:atmoeq}
\end{figure}

To find the best-fit model for both our nightside and dayside brown dwarf spectra, we multiply the ATMO 2020 models by the scale factor $\Big( \frac{R_{\text{BD}}}{D} \Big)^2$, where the brown dwarf radius, $R_{\text{BD}}$, and distance, $D$, are those from \cite{wd1032}. We then perform a $\chi^2$ fitting between the model and our data, as we did in Section \ref{sec:fieldbds}. We take the smallest $\chi^2$ value to be our best fitting model. For the chemical equilibrium grid, the best fitting models are shown in upper and lower panels of Figure \ref{fig:atmoeq} for the dayside and nightside of \objb{}, respectively.

For the nightside of \objb, the best-fitting ATMO 2020 model has $T_{\text{eff}} = 1800$~K and log $g$ = 3.0 compared to $T_{\text{eff}} = 2100$~K and log $g$ = 2.5 for the best-fitting dayside model. These results are consistent with our brightness temperature calculations. They also imply a potential low gravity for \objb. To evaluate the surface gravity of our brown dwarf, we calculated the gravity indices from \cite{gravind}, however these were inconclusive due to the limited wavelength range of our data. We calculate the surface gravity of \obj{} from its mass and radius, derived by \cite{wd1032} via radial velocity and eclipse photometry respectively, as log $g$ = $5.21 \pm 0.09$, meaning that the ATMO models fit uncharacteristically low surface gravities. It should be noted that the disequilibrium grid of ATMO models produce a more reasonable surface gravity of log $g$ = 5.5 for the dayside of the brown dwarf, however they do not fit the spectral features, and still underestimate the nightside as log $g$ = 3.0, which is unphysical given the mass and radius of \objb{}. Additionally, even if we fix the surface gravity to only vary between 5.0--5.5, the best-fit ATMO models do not fit the water absorption feature or the rest of the spectra morphology well for either the dayside or the nightside. 

We also fit the Sonora suite of models, which are designed for non-irradiated sub-stellar objects, to our dayside and nightside spectra \cite{sonora}. We find that they give similar results to the ATMO models, however they produce more realistic surface gravity estimates of log $g=5.5$. Additionally, the low metallicity grid of Sonora models provides the best fits to our data, particularly on the dayside, indicating that \objb{} may be metal-poor. Overall, the ATMO models fit our data better despite not matching the surface gravity, with $\chi^2$ values consistently less than half of those produced by the Sonora models, which is likely a consequence of better matching the pressure-temperature profile, the cloud parameters, and chemical abundances. Both the ATMO and Sonora models do not fit the water absorption feature or the slope shortwards of 13000~\AA{} particularly well, indicating that these non-irradiated models do not adequately describe our observations, proving the effects of irradiation need to be included.

\subsection{Forward Models}
\label{subsec:forwardmodels}

We run a small grid of 1D radiative-convective equilibrium models using EGP \citep{marley1999, ackerman2001}. The grid spans the following parameters: Age = 1, 1.5, and 2~Gyr , Metallicity = $-1\times$, $-0.5\times$ and 0$\times$ solar, and the cloud cases of Cloud Free, Cloudy with sedimentation efficiency, f$_{sed}$ = 3.0 or 5.0 \citep{ackerman2001}. Available clouds for condensation include KCl, ZnS, Na$_2$S, MnS, Cr, MgSiO$_3$, Fe, and Al$_2$O$_3$.

By varying the recirculation factor, which parameterises the redistribution of incident energy across the atmosphere, we can model both the day and nightsides. The age is used to set the internal heat flux, $T_\mathrm{int}$, from the \cite{marley2018} evolution grid. These models assume chemical equilibrium. We do not include TiO and VO opacity in the atmosphere. Given the hot temperature of the white dwarf we employ the same methodology as \cite{rachael} and \cite{ben}, where we increase the irradiation in the first few wavelength bins to account for the flux at wavelengths shorter than the modeling grid, essentially assuming that the opacities are constant through the ultraviolet where opacities are not available.

\begin{figure}
    \centering    
    \includegraphics[width=\columnwidth]{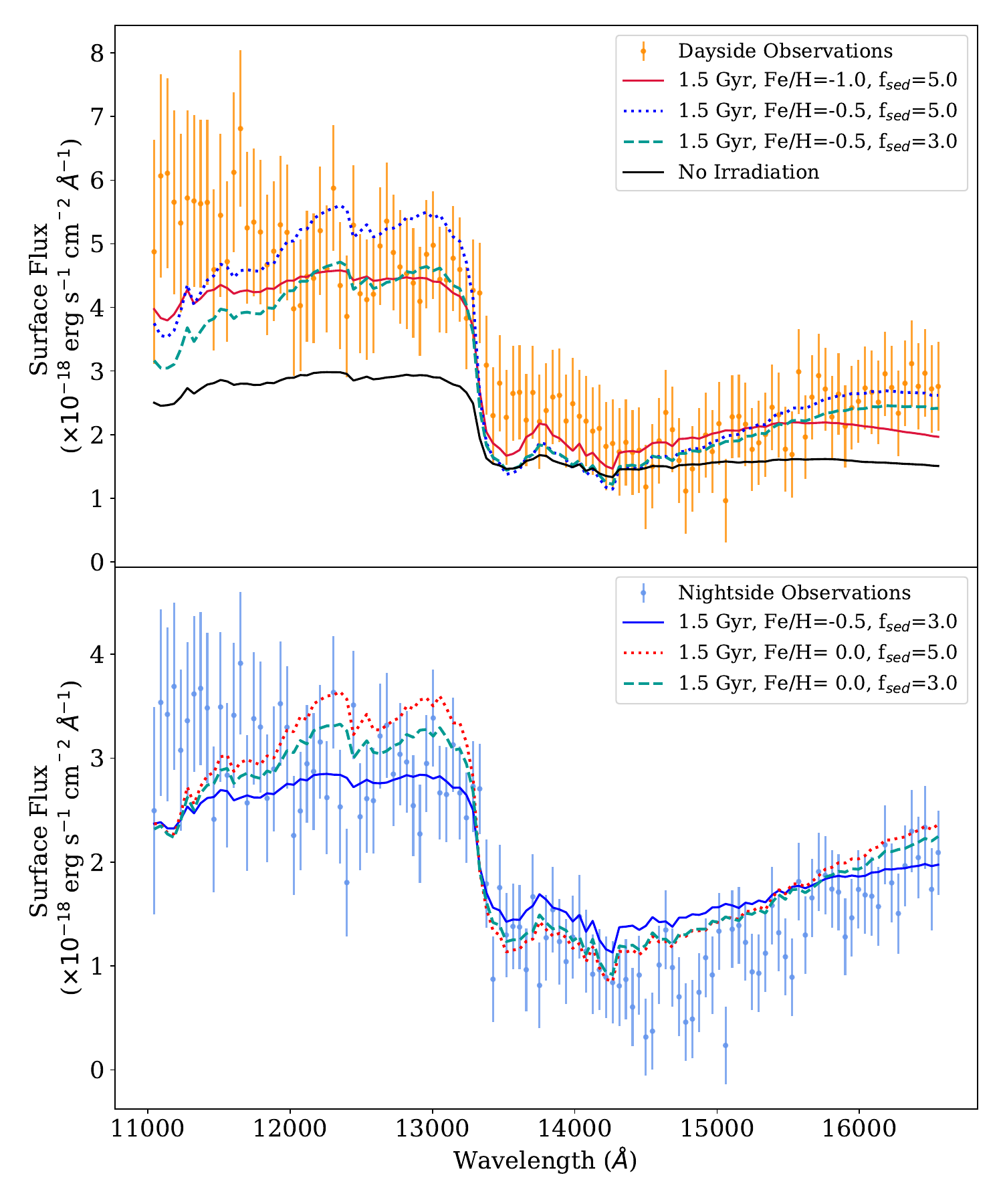}
    \caption{Best fit forward models of \objb{}. The dayside spectrum is shown in the upper panel in orange with the three best-fit models depicted by a red solid line, a blue dotted line and a green dashed line respectively. The black solid line shows the best-fit model with the effects of irradiation removed. The nightside spectrum is shown in the lower panel in light blue with the three best-fit models depicted by a blue solid line, red dotted line and green dashed line respectively. The age, metallicity and cloud parameter of each model are shown in the legend. An age of 1.5~Gyr corresponds to an internal heat flux of $\sim$1775~K.}
    \label{fig:forwardmodels}
\end{figure}

\begin{figure}
    \centering      
    \includegraphics[width=\columnwidth]{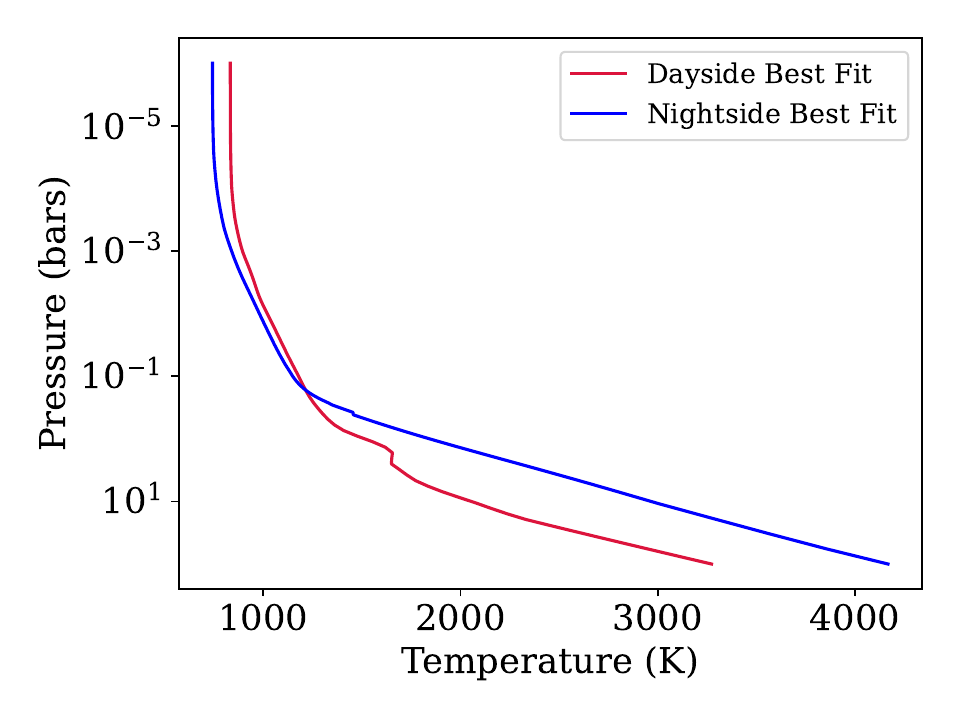}
    \caption{Pressure-temperature profiles of the best fit PICASO forward models of \objb{}. The dayside pressure-temperature profile is shown in red, and the nightside pressure-temperature profile is shown in blue.}
    \label{fig:tplaura}
\end{figure}

Using PICASO \citep{batalha2019}, we compute the resultant spectra and select the best-fits to the data based on the chi-squared metric. The best-fit forward models for the data are shown in Figure \ref{fig:forwardmodels}, with the dayside shown in the upper panel and the nightside shown in the lower panel. These models include the reflected flux for the dayside. The best-fit model has an age of 1.5~Gyr, which corresponds to an internal heat flux of $\sim$1775~K, with Fe/H = -1.0 and f$_{sed}$ = 5.0. Without the effects of irradiation the model fits poorly, showing that irradiation is present in the atmosphere and needs to be considered. The associated pressure-temperature profiles are shown in Figure \ref{fig:tplaura}. These pressure-temperature profiles do not show a temperature inversion on either the dayside or the nightside. The models predict a Na and K feature in the 1.1--1.2~$\upmu$m wavelength region which is not supported by the data. Thus with the removal of those features, better fits are achieved overall but the models still struggle to reproduce the slope in the blue end of the spectrum.

\subsection{Atmospheric Retrievals}
\label{subsec:irrbdmodels}

To fit the phase-resolved dayside and nightside spectra with more flexibility than the self-consistent grids, we also used the PETRA retrieval framework \citep{petra}, which uses the PHOENIX atmosphere model \citep{Hauschildt1999, Barman2001} as the forward model in a Different Evolution Markov Chain Monte Carlo statistical framework \citep{TerBraak2006}. The retrievals were run on a 64-layer pressure grid varying the surface gravity, the pressure-level and opacity of a grey cloud-deck, and the temperature structure. To parameterise the temperature structure, we used the 5-parameter \cite{ParmGuill2014} parameterisation, with the internal temperature as an additional free parameter. The composition was kept at solar metallicity, with the most important opacity sources being H$_2$O \citep{Barber2006}, H- \citep{John1988}, and H$_2$-H$_2$ and H$_2$-He collision induced absorption (CIA, \citealt{Borysow1989, Borysow1990}).

\begin{figure}
    \centering    
    \includegraphics[width=\columnwidth]{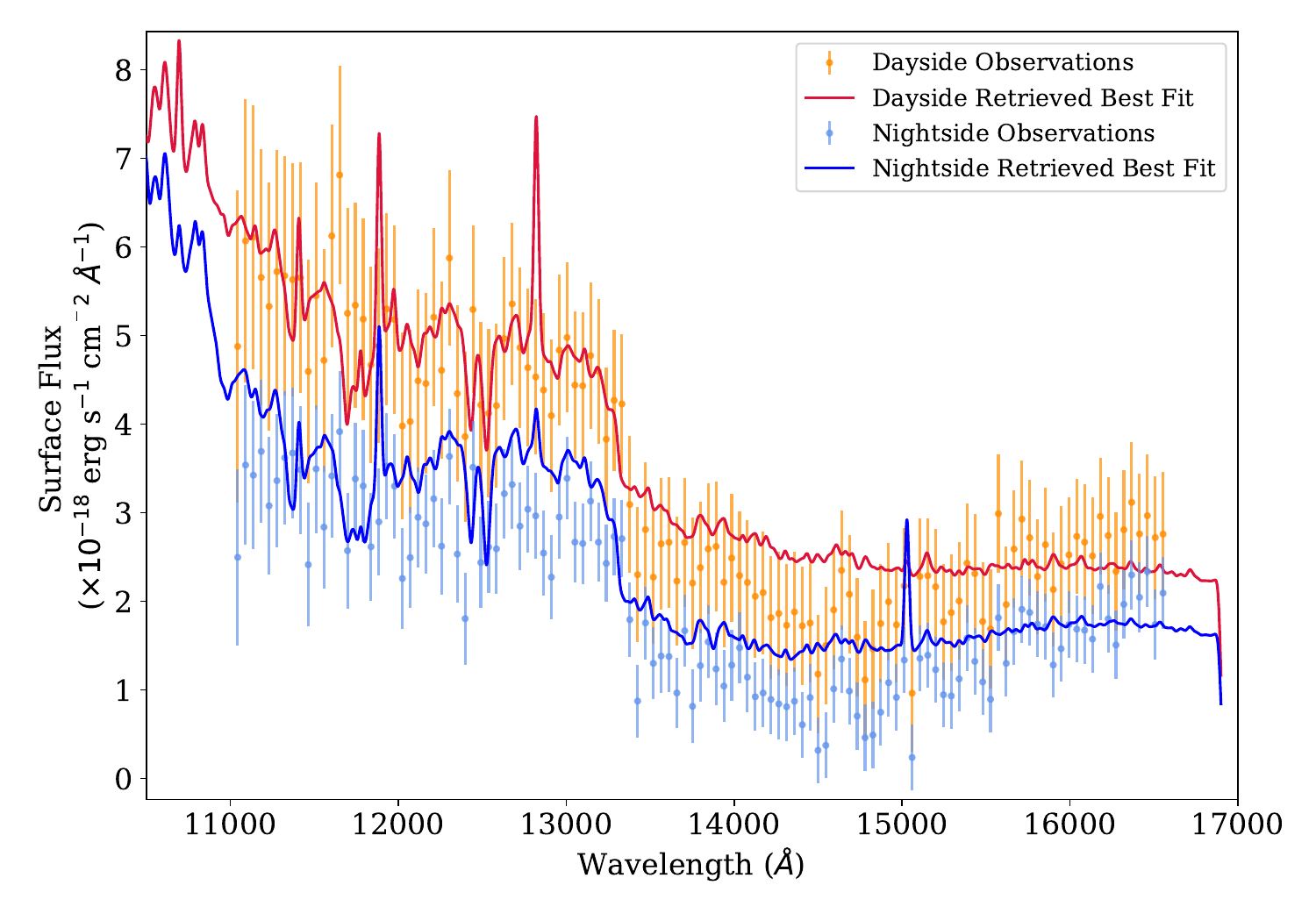}
    \caption{Best fit retrieved spectra of \objb{} using PETRA. The dayside observations are shown in orange with the fit overlaid in red. The nightside observations are shown in light blue with the fit overlaid in dark blue.}
    \label{fig:retrieval_bestfits}
\end{figure}

\begin{figure}
    \centering      \includegraphics[width=\columnwidth]{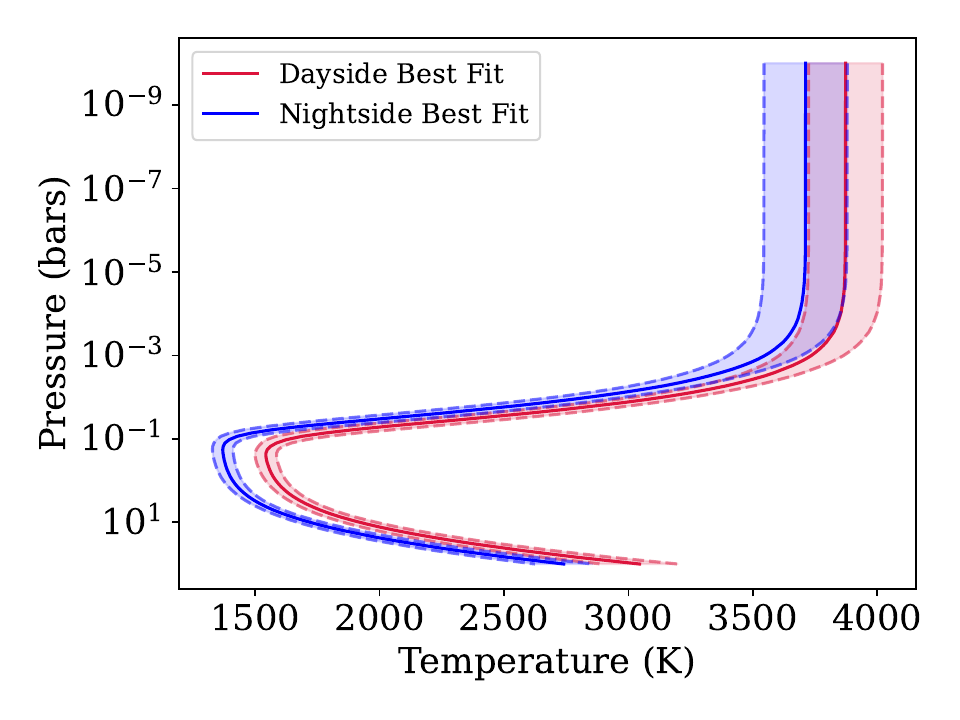}
    \caption{Pressure-temperature profiles of the best fit retrieved spectra of \objb{} using PETRA. The dayside pressure-temperature profile is shown in red, and the nightside pressure-temperature profile is shown in blue. The filled regions between the dashed lines indicate 1$\sigma$ confidence limits, in red and blue for the dayside and nightside respectively.}
    \label{fig:tpjosh}
\end{figure}

Figure~\ref{fig:retrieval_bestfits} shows the best-fit dayside and nightside spectra. The associated pressure-temperature profiles are shown in Figure \ref{fig:tpjosh}. Both the dayside and nightside retrievals preferred a strongly inverted atmosphere. An irradiation-driven temperature inversion could be expected on the dayside, as seen in other irradiated brown dwarfs (e.g. WD0137B, \citealt{Lee2020}). However, such an inversion would not be expected to be retained on the nightside, as the photosphere would have more than enough time to have radiatively cooled to a non-inverted profile. At a pressure of 1~bar, the radiative timescale in the photosphere is 0.183~hours, which is equal to 8.3\% of a full orbit of \objab{}. This timescale decreases at lower pressures, so the atmosphere will have cooled to a non-inverted profile well within the 1.1~hours between our observations of the dayside and nightside. 

\begin{figure}
    \centering    
    \includegraphics[width=\columnwidth]{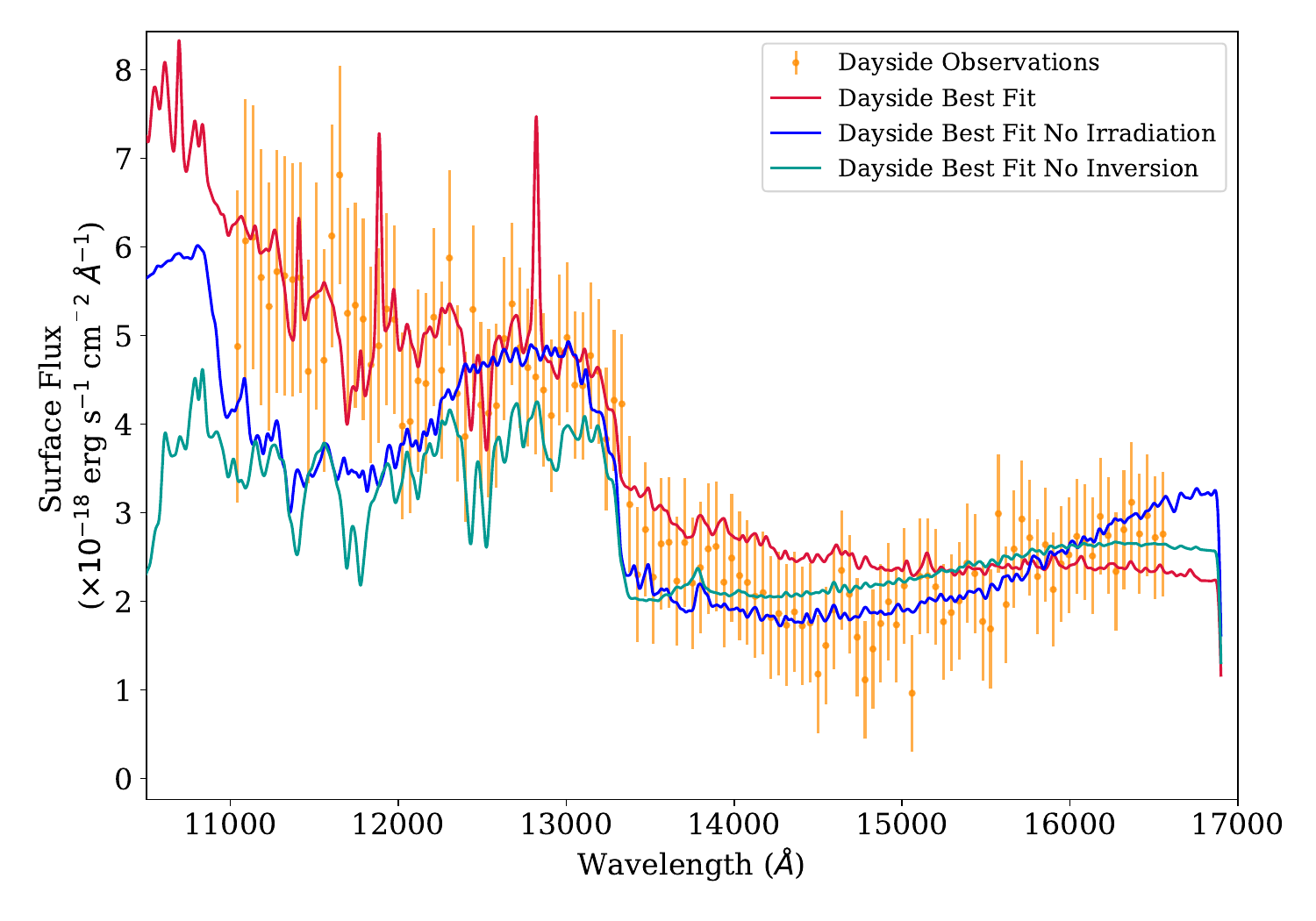}
    \caption{Best fit retrieved spectra of the dayside of \objb{} considering the effects of irradiation and inversion. The dayside spectrum is shown in orange. The red line shows the best fit, which includes both irradiation and a temperature inversion. The blue line is the best fit without irradiation, and the green line is the best fit without a temperature inversion. The red, blue and green lines are shown from top to bottom respectively at the short wavelength end.}
    \label{fig:retrieval_dayside}
\end{figure}

Figure~\ref{fig:retrieval_dayside} shows the dayside best-fit spectra with and without the inversion, as well as without the irradiation (i.e., just T$_{int}$ determining the temperature structure in the Parmentier \& Guillot parameterisation). As can be seen, the retrieval adds the inversion to increase the flux between 1.1 and 1.3~$\upmu$m, while sacrificing the fit longward of 1.4~$\upmu$m. Without either the inversion or irradiation, the retrieval struggles to match the short-wavelength data. 

While the nightside temperature inversion is likely unphysical as it is driven by irradiation, both retrievals find a reasonable surface gravity, retrieving log$_{10}$(g$_{cgs}$) = 5.57 $\pm$ 0.22 and 5.22 $\pm$ 0.22 from the dayside and nightside respectively. The retrieved temperatures are 1748 $^{+76}_{-62}$ and 1555 $^{+66}_{-67}$K, respectively. The true internal temperature is expected to be the same between the dayside and nightside, so the increased temperature on the dayside represents the contribution from the irradiation that makes it to the deep atmosphere. Neither the dayside nor the nightside retrieved a cloud at observable photospheric pressures, though the PICASO forward models indicate that clouds may help fit the short wavelength observations more physically than a temperature inversion. Since the irradiated forward model is not showing such a large inversion, the processes responsible for the retrieved inversion need to be further explored. Absorption by unmodelled photochemical products including disequilibrium gases and hazes as well as energy transport by atmospheric waves are among the possibilities (e.g. \citealt{inversionposs}).

\section{Discussion}
\label{sec:discussion}
With comprehensive system parameters from \cite{wd1032} (see Table \ref{table:wd1032params}) resulting from high-resolution photometry and multiple spectroscopic observations, we do not recalculate those system parameters in this paper. When conducting our MCMC analysis on the broadband and sub-band lightcurves we derived for \obj, we treated the period in Equation \ref{eq:mcmcmodel} as a fixed parameter with a value of $P = 0.09155899610$~days. Fixing the period to this value yielded excellent agreement between our data and the best-fit MCMC model. We note that when the period was allowed to vary as a free parameter in our analysis, the true period was successfully recovered within 1$\sigma$. In addition, when the period was allowed to vary whilst fitting the MCMC model to our sub-band lightcurves, the individual periods for the $J$-, water and $H$-band data were all within 1$\sigma$ of the true period and each other.

\subsection{Comparison to Field Brown Dwarfs}
\label{subsec:fieldbdcomp}

To evaluate the effects that the external irradiation from the white dwarf has on the spectrum of the irradiated brown dwarf, we compare the dayside and nightside spectra of \objb{} to field brown dwarfs. Most field brown dwarfs exhibit rotational modulations in their emission spectra  \citep{Buenzli2014, cloudrot}. These flux variations arise from the cloud thickness variations in the atmosphere of the brown dwarfs \citep{apai, hetclouds}. The cloud structure visible to external observers changes dramatically at the L/T spectral type transition \citep{Radigan2012}. At these temperatures, the cloud top is modulated by zonal circulation and atmospheric waves \citep[][]{Apai2017, Apai2021, Zhou2022, Fuda2024, Plummer2024}. At temperatures below, the cloud top is thought to sink below the photosphere and thus is  no longer visible to the observer \citep{cloudsink}. In irradiated and tidally locked brown dwarfs, rotational phase-dependent changes in flux can arise from the constant irradiation on the dayside and atmospheric circulation from the dayside to the nightside.

We searched the SpeX prism and Cloud Atlas databases which contain spectra of L, T and Y dwarfs for the field brown dwarf that best-matched \objb. We normalised the spectra such that the flux at 13000~\AA{} is at a value of 1. We find that the Cloud Atlas spectral library does not extend to early enough spectral types, whereas the SpeX prism spectral library contains numerous early L-dwarfs. Our nightside spectrum is best fit by L1 peculiar brown dwarfs, which exhibit deeper water absorption and a bluer slope longwards of 13000~\AA{} compared to their L1 counterparts \citep{L1pec}. Peculiar brown dwarfs often exhibit higher or lower metallicities than field brown dwarfs \citep{BDpecmetal}, and the differences in their spectra may result from changes in cloud structure and opacity alongside atmospheric changes. A spectral type of L1 pec is also consistent with our average nightside brightness temperature of 1909~K. We note that \objb{} was originally posited as an L5 spectral type by \cite{wd1032} however, the GNIRS spectrum they analysed was normalised to photometry that was taken at different points in orbital phase. The UKIDSS $Y$ and $J$ photometry was taken separately to the $H$ and $K$ photometry. If the $H$ and $K$ photometry were taken during or close to the eclipse, then their relative magnitude is lower than it should be compared to the $Y$ and $J$ photometry. This would then result in a later spectral type being identified as the best fit, where an earlier spectral type is actually more appropriate.

The dayside of \objb{} however, is instead best fit by a subdwarf sdL0 object, WISE~J04592121+1540592 \citep{sdL0}. Subdwarfs describe metal-poor objects with subsolar abundances that often exhibit blue colours in the near-infrared and enhanced absorption bands for metal hydrides, but otherwise resemble the morphology of M, L, T and Y dwarf spectra \citep{subdwarf2, subdwarf1}. The kinematics of the white dwarf correspond to a thick disc membership, meaning that \objb{} could be metal-poor due to its age \citep{wd1032}. However, the subdwarf field brown dwarf spectra do not fit the shape of the water absorption feature centred on $\sim$14000~\AA{} well. Subdwarfs tend to be less cloudy or completely cloud-free due to the reduction of condensates in their atmospheres, which causes their low metallicity \citep{sdclouds}. Clouds are typically expected in early L-type brown dwarfs as this is before the L/T transition where condensate clouds sink below the photosphere \citep{lttransition}. As the dayside of \objb{} is best-matched by subdwarf spectra, which appear flatter in the $J$-band, it is likely that the irradiation from the white dwarf is affecting the condensate clouds in such a way that the atmosphere of the brown dwarf is appearing more metal-poor. This suggests that the strong irradiation of the dayside is either dissipating some of the cloud coverage due to the increased heat, pushing silicate clouds below the photosphere, or circulating clouds towards the nightside, so that the dayside is silicate cloud-free \citep{burrows}. This causes a heterogeneous atmospheric structure that is observed in the phase-dependent spectra of the brown dwarf \citep{cloudsirrad}.

\begin{figure}
    \centering    
    \includegraphics[width=\columnwidth]{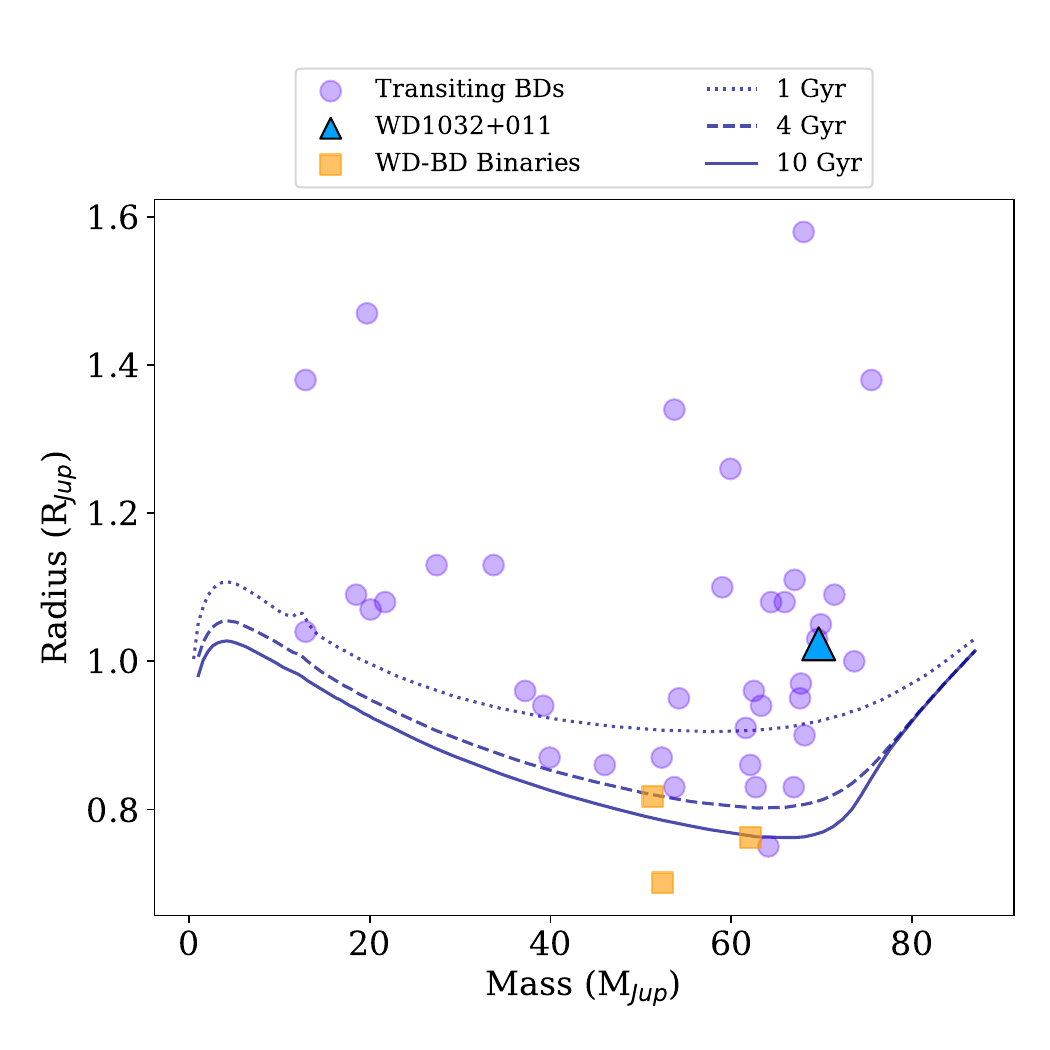}
    \caption{Known transiting brown dwarfs in the mass--radius parameter space, represented by purple circles alongside eclipsing brown dwarf companions to white dwarfs represented by orange squares. \objb{} is shown with an outlined blue triangle. The dotted, dashed and solid navy lines are Sonora brown dwarf evolution models for 1~Gyr, 4~Gyr and 10~Gyr respectively. The sample of transiting brown dwarfs is taken from \protect\cite{beth}. The transiting exoplanets are taken from the NASA Exoplanet Archive.}
    \label{fig:transBDmassradius}
\end{figure}

Since \obj{} is an eclipsing system, \objb{} belongs in the small population of transiting brown dwarfs that have been discovered. We compare \objb{} to this population of transiting brown dwarfs around main sequence stars, which was most recently updated by \cite{beth}. We also consider the other 3 known eclipsing close white dwarf--brown dwarf binaries, SDSS~J1411+2009 (hereafter SDSS1411, \citealt{ben}), SDSS~J1205-0242 \citep{J1205}, and ZTF~J0038+2030 \citep{ztf}. In Figure \ref{fig:transBDmassradius} we place \objb{} in the mass--radius parameter space of known transiting brown dwarfs. The purple circles show the transiting brown dwarfs around main sequence stars, the orange squares correspond to eclipsing brown dwarfs orbiting white dwarfs, and \objb{} is represented by the outlined blue triangle. As can be seen, the brown dwarfs in eclipsing white dwarf--brown dwarf binaries fit well within the population of higher mass transiting brown dwarfs, alongside the other brown dwarfs orbiting white dwarfs. SDSS1411B has a radius slightly lower than the majority of the transiting brown dwarfs due to it being a later spectral type of T5, with brown dwarfs continuing to contract whilst they age and cool \citep{ben}. The other brown dwarf companions to white dwarfs are well-aligned in the parameter space, with masses and radii consistent with the expectations from brown dwarf evolutionary models. However, \objb{} has a higher radius than expected compared to both the Sonora evolution models and the other eclipsing brown dwarf companions to white dwarfs, verifying that it is inflated.

\begin{figure}
    \centering    
    \includegraphics[width=\columnwidth]{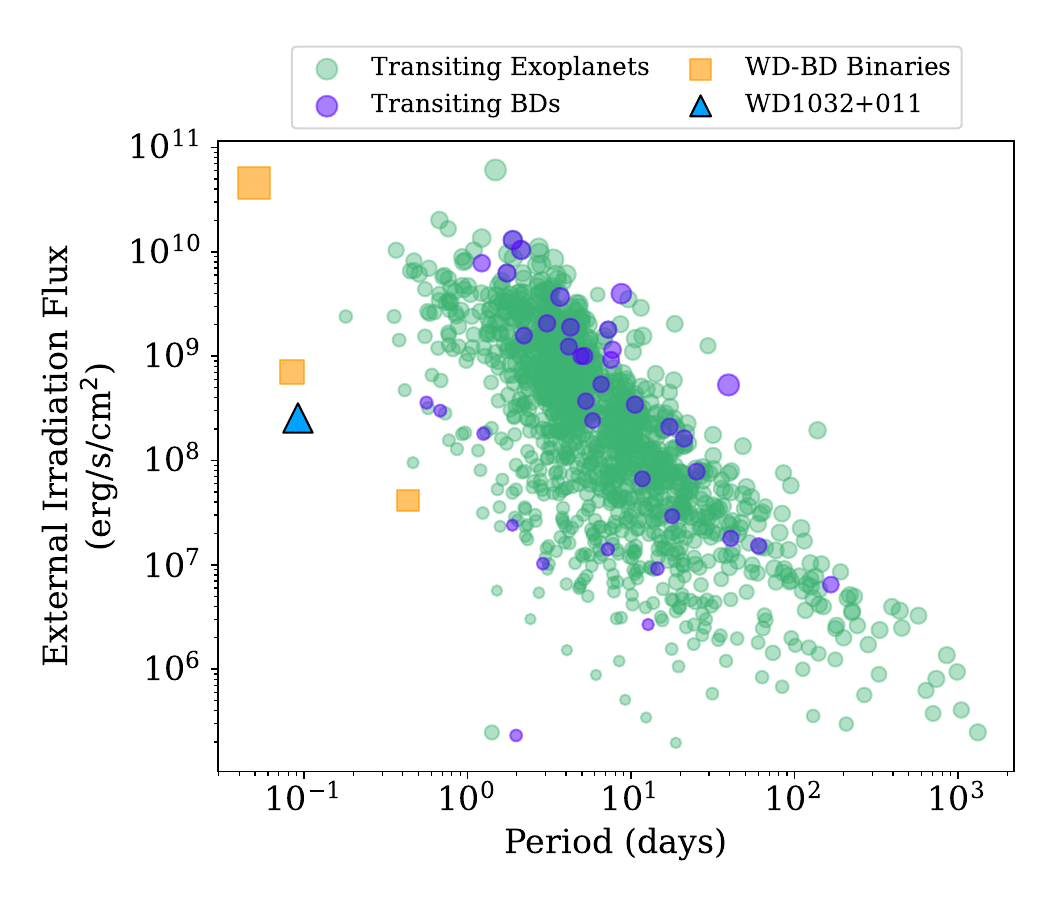}
    \caption{Variation of external irradiation flux received by brown dwarfs and exoplanets with orbital period. The purple circles correspond to transiting brown dwarfs, and the green circles are transiting exoplanets. The orange squares show eclipsing brown dwarf companions to white dwarfs. \objb{} is represented by the outlined blue triangle. The size of the points corresponds to the effective temperature of the host star. The sample of transiting brown dwarfs is taken from \protect\cite{beth}. The transiting exoplanets are taken from the NASA Exoplanet Archive.}
    \label{fig:transBDirrperiod}
\end{figure}

In Figure \ref{fig:transBDirrperiod}, we investigate the variation of external irradiation flux with period for transiting brown dwarfs and exoplanets, alongside the population of brown dwarfs in close orbits around white dwarfs. Although \objb{} and the other brown dwarfs irradiated by white dwarfs follow the general trend of the transiting brown dwarfs, they should have an increased external irradiation flux considering their periods to continue the same linear trend seen in the non-irradiated transiting brown dwarfs. 

Since white dwarfs are much smaller than main sequence stars, their brown dwarf companions would receive a lower external irradiation flux than if they were orbiting a main sequence star at the same temperature, as the decreased radius of the white dwarf decreases its surface area and thus the external irradiation flux received by the brown dwarf \citep{wdhr}. Additionally, as the white dwarfs are already evolved, the orbits of the brown dwarfs have shrunk during this evolution, allowing them to occupy orbits much closer to their primary stars than the brown dwarfs transiting main sequence stars. This discrepancy leads to two distinct populations in Figure \ref{fig:transBDirrperiod}, separating the highly irradiated brown dwarfs from the other transiting brown dwarfs. Both of these populations follow slightly different linear trends as decreasing the orbital period increases the irradiation flux received by the brown dwarf.

\subsection{Brown Dwarf Atmosphere Models}

To quantify the differences between the dayside and nightside of \objb{} and the effects of irradiation on the atmosphere, we analysed one dimensional atmospheric models for both hemispheres. We compared our derived dayside and nightside spectra to the ATMO 2020 atmospheric models which are non-irradiated, generated PICASO forward models which consider irradiation, and ran retrievals using PHOENIX atmosphere models with PETRA. 
From all of these models, we found that there is a consistent temperature contrast between the dayside and nightside of \objb{}, which is due to irradiation coupled with a poor heat redistribution between the hemispheres. The best-fit ATMO models for both the dayside and the nightside have effective temperatures that agree with our brightness temperatures (Figure \ref{fig:brighttemp}), and they have low gravities of log~$g =$~2.5--3.0. However, given the mass and radius of \objb{}, we calculate its surface gravity as log~$g = 5.21$. The uncharacteristically low gravity identified by fitting to the ATMO models is likely because they are non-irradiated models, and a lower surface gravity better replicates some of the features seen in the irradiated spectra.

The best-fit PICASO forward models for both the dayside and nightside have ages of 1.5~Gyr, which correspond to an internal heat flux of $\sim$1775~K. The best-fit models also favour low metallicity, which corroborates the spectral type of L1 peculiar and the metal-poor spectral features evidenced in Section \ref{subsec:fieldbdcomp}. The PICASO models do not simultaneously fit both the short wavelength data and the water absorption feature well, however the presence of clouds improves the fit to the short wavelength data. The models predict a Na and K feature in the 1.1--1.2~$\upmu$m wavelength region which is not supported by the data. Na and K are both easily ionised, and the fact that we do not see evidence of their predicted features indicates that the ultraviolet heating from the white dwarf is impacting these species.

The retrievals performed on the PHOENIX model grid, which accounts for irradiation, match both the dayside and nightside spectra well. An irradiation-driven temperature inversion is seen in the dayside, a feature which is often seen in the atmospheres of irradiated brown dwarfs and exoplanets, arising due to the strong absorption of incoming ultraviolet flux by molecules such as VO and TiO \citep{inversion1, inversion2}. There is an unphysical temperature inversion seen in the nightside spectrum, but this is likely due to the increase in flux shortwards of 13000~\AA{}. Such an inversion would not be expected to be retained on the nightside, as the photosphere would have more than enough time to have radiatively cooled to a non-inverted profile. For WD0137B, an irradiated brown dwarf which has a higher external irradiation flux with a similar orbital period of 114~min compared to the 132~min orbital period of \objb{}, the atmosphere relaxes to a non-inverted state on the nightside where the temperature is no longer externally forced by irradiation \citep{0137invert}. Therefore, \objb{} is expected to have adequate time to radiatively cool to a non-inverted temperature profile on the nightside, which also indicates that the radiative timescale it takes to cool is significantly shorter than the advective timescale at which heat transport occurs between the hemispheres. The slope present in the short wavelength data could arise from the low metallicity of the brown dwarf and its L1 peculiar spectral type. However, as this slope is not adequately fit by the retrievals, the fit introduces an unphysical temperature inversion on the nightside, which increases the flux bluewards of 13000~\AA{} and thus produces a better fit to the data.  The slope could also arise from clouds, but the atmospheric retrievals do not recover clouds above the photosphere. Future work with more complex retrieval analyses, or a simultaneous white dwarf--brown dwarf retrieval framework may improve the retrievals and help explore why an unphysical temperature inversion is chosen to fit the spectra.

The irradiated retrievals recover surface gravities within 2$\sigma$ of the calculated value for the dayside, and within 1$\sigma$ for the nightside. The internal temperatures retrieved corroborate the dayside-nightside temperature contrast derived from brightness temperature, with a 193~K temperature difference between the hemispheres. The performance of these irradiated model retrievals show that \objb{} has a dayside-nightside temperature contrast that is driven by the irradiation received by the white dwarf, and that this irradiation is altering the atmospheric profile of the brown dwarf.

\subsection{Comparison to Hot Jupiters}

In field brown dwarfs, the effective temperature is solely dependent on the internal heat flux of the brown dwarf, which cools as it evolves through its lifetime. However, for irradiated brown dwarfs, the constant external irradiation flux received from the host star increases their effective temperature such that 

\begin{equation}
~~~~~~~~~~~~~~~~~~~~~~~~~~~~T_{\text{eff}} = (T_{internal}^4 + T_{irradiation}^4)^{\frac{1}{4}}.
\end{equation}

Short-period irradiated brown dwarfs companions to white dwarfs have equilibrium temperatures which are comparable to the hottest hot Jupiters, making them excellent proxies for studying hot Jupiters. In particular, the observed spectra of hot Jupiters are influenced by the presence and composition of clouds, which alter their atmospheric pressure-temperature profiles, with silicate clouds potentially disappearing for equilibrium temperatures cooler than 1600~K \citep{hjpt, hjclouds}. Irradiated brown dwarfs, particularly higher contrast objects around white dwarfs, enable us to gain insight into the connections between non-irradiated brown dwarfs and hot Jupiters, and how clouds may influence their atmospheric structure and temperature profiles.

\begin{figure}
    \centering    
    \includegraphics[width=\columnwidth]{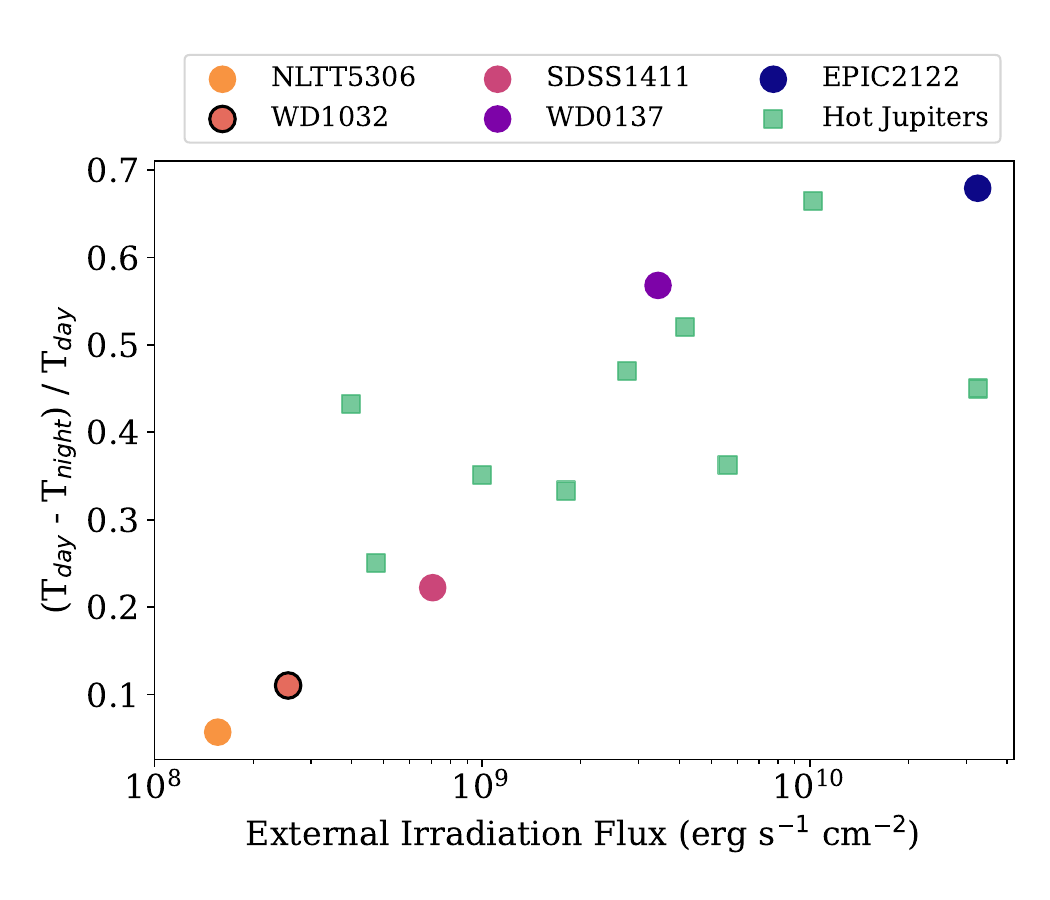}
    \caption{Comparison of dayside-nightside temperature contrast and external irradiation flux for \protect\objb{} alongside the irradiated brown dwarf companions to white dwarfs NLTT5306B, SDSS1411B, WD0137B and EPIC2122B, all shown as circles. Hot Jupiters are shown with green squares, with systems selected from \protect\cite{Beatty} and \protect\cite{Komacek}.}
    \label{fig:HJWDBD}
\end{figure}

To compare \objb{} to irradiated brown dwarfs and hot Jupiters, we calculated the external irradiation flux received from the white dwarf, as well as its dayside-nightside temperature contrast relative to the dayside temperature. We compare \objb{} to the sample of 4 irradiated brown dwarf companions to white dwarfs alongside a selection of hot Jupiters in Figure \ref{fig:HJWDBD}. The irradiated brown dwarfs are NLTT5306B (L5, P=101.88~min; \citealt{rachael}), SDSS1411B (T5, P=2.02864~hr; \citealt{ben}), WD0137B (L6--L8, P=116~min; \citealt{yifan}) and EPIC2122B (L3, P=68.21~min; \citealt{epic, yifan}). The hot Jupiters included in this comparison are selected from the samples analysed by \cite{Beatty} and \cite{Komacek}. These consist of HD~149026b \citep{149026}, HD~189733b \citep{knutson1, knutson2, knutson3}, HD~209458b \citep{crossfield, zellem}, HAT-P-7b \citep{Wong2016}, WASP-12b \citep{cowan}, WASP-14b \citep{wong2}, WASP-19b \citep{Wong2016}, WASP-33b \citep{zhang, chakra}, WASP-43b \citep{stevenson, kokori, bell} and WASP-103b \citep{kreidberg}.

The irradiated brown dwarfs in Figure \ref{fig:HJWDBD} are represented by circles, with each system labelled. The hot Jupiters are represented by green squares. We report the relative dayside-nightside temperature contrast relative to the dayside temperature, that is $(T_{day} - T_{night})/T_{day}$. Using this metric allows direct comparison between the dayside-nightside temperature contrasts of irradiated brown dwarfs and hot Jupiters, whereas the difference between the dayside and nightside temperatures alone may be sensitive to the different instruments used for observations. As can be seen, as external irradiation flux increases, often caused by a higher stellar temperature or a smaller orbital separation, the dayside-nightside temperature contrast metric also increases. \cite{showguill} predicted that hot Jupiters would follow this trend, and the irradiated brown dwarfs considered here are interspersed with the hot Jupiters and follow this trend as well. This shows that irradiated brown dwarfs can be effectively utilised as proxies for hot Jupiters, and indicates that they may undergo similar cloud and atmospheric changes due to their irradiation. \objb{} is well-aligned with both irradiated brown dwarfs and hot Jupiters. It receives less external irradation flux than the majority of other known irradiated brown dwarfs, however this flux is still high enough to cause a moderate dayside-nightside temperature contrast and cause atmospheric changes on the dayside.

\subsection{Comparison to Irradiated White Dwarf--Brown Dwarf Binaries}

Irradiated brown dwarfs reside between hot Jupiters and non-irradiated brown dwarfs, and their atmospheres can thus be used to investigate key atmospheric processes such as condensate cloud formation and dissipation, and heat redistribution from the dayside to the nightside. \objb{} is the fifth irradiated brown dwarf companion to a white dwarf which has been studied by high-precision, time-resolved HST/WFC3 spectrophotometry, following SDSS1411B, WD0137B, EPIC2122B and NLTT5306B \citep[][]{ben, yifan, rachael}. \objb{} sits between NLTT5306B and the other highly irradiated white dwarf--brown dwarf binaries, as can be seen in Figure \ref{fig:HJWDBD}. Similarly, the dayside-nightside temperature contrast metric is between those for NLTT5306B and SDSS1411B, WD0137B and EPIC2122B. Dayside-nightside temperature contrast is influenced by the rotation rate of the brown dwarf, which is faster for shorter period brown dwarfs \citep{TanShow2020}. With the exception of NLTT5306B, these irradiated brown dwarfs all follow that trend with \objb{} having a period of 2.2~hours, SDSS1411B has a period of 2.0~hours, WD0137B with an orbit of 1.9~hours and EPIC2122B having a period of only 68.21~min. EPIC2122B receives 127 times more external irradiation flux than \objb{}, whereas WD0137B, SDSS1411B and NLTT5306B receive 13.7 times more, 2.8 times more and 1.6 times less respectively. All of these irradiated brown dwarfs follow the trend of increasing dayside-nightside temperature contrast as external irradiation flux also increases, which is also observed in hot Jupiters (e.g. \citealt{Komacek, Beatty}).

\begin{figure}
    \centering    
    \includegraphics[width=1.1
\columnwidth]{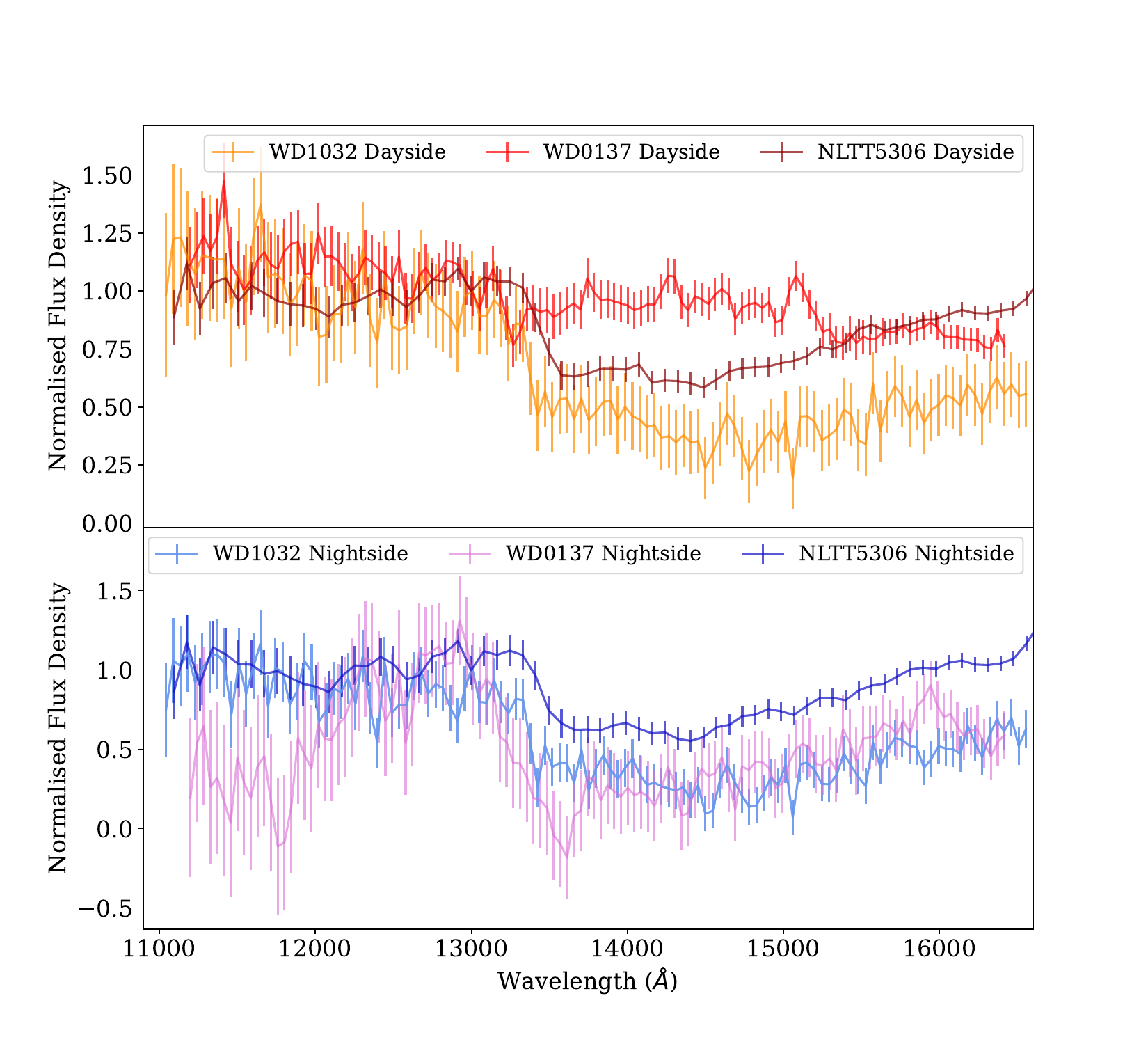}
    \caption{Comparison of dayside and nightside spectra of \objb{} alongside WD0137B \protect\citep{yifan} and NLTT5306B \protect\citep{rachael}. The upper panel shows the dayside spectra of these objects and the lower panel shows the nightside spectra. Spectra of \objb{} are in orange and light blue, WD0137B is shown in red and purple, and NLTT5306B is shown in dark red and dark blue. All spectra have been normalised such that the flux at 13000~\AA{} is at a value of 1.}
    \label{fig:WDBDDayNight}
\end{figure}

We compare the dayside and nightside spectra of \objb{} with the non-eclipsing systems NLTT5306B and WD0137B as these have similar spectral types and differing levels of irradiation. We do not compare to EPIC2122B due to the difference in external irradiation flux being too vast to show the key effects of irradiation. Similarly, we do not compare to the eclipsing system SDSS1411B due to its significantly later spectral type, which causes morphological differences in its spectra. Figure \ref{fig:WDBDDayNight} shows the dayside and nightside spectra of these three irradiated brown dwarfs, with all spectra normalised such that the flux at 13000~\AA{} is at a value of 1.

As can be seen, the dayside and nightside spectra for \objb{} and NLTT5306B are particularly similar, following the same overall trend with only minor differences in the depth of the water feature. WD0137B exhibits a much flatter dayside spectrum and a nightside spectrum that differs shortwards of 13000~\AA{}. As WD0137B receives an external irradiation flux an order of magnitude higher than that received by \objb{} and NLTT5306B, this apparent flattening of the water absorption feature present in the spectrum of the dayside is likely due to an increased level of irradiation. WD0137B also exhibits an irradiation-driven temperature inversion on the dayside \citep{Lee2020}. Additionally, the nightside spectrum of WD0137B does not exhibit the the rise in flux seen in at shorter wavelengths in the spectra of both \objb{} and NLTT5306B, which deviates from what would be expected from atmospheric models. Part of this difference in the $J$-band of \objb{} can be explained by the best-fitting spectral type being an L1 peculiar type, which tend to exhibit this slope bluewards of 13000~\AA{}. However, this characteristic is also seen for NLTT5306B, which has been classified as an L5 spectral type. We note that both of these objects are inflated, whereas WD0137B does not exhibit signs of inflation \citep{ben}. Thus, it seems that the inflation of the brown dwarf, which is likely due to the irradiation from the white dwarf slowing down the contraction of the brown dwarf, is causing this increase at the short wavelength end of the dayside and nightside spectra of both \objb{} and NLTT5306B.

\subsection{Inflation of \objb{}}
As they age, brown dwarfs cool, contract, and evolve through spectral types M, L, T and Y. Less massive brown dwarfs tend to have a faster cooling rate compared to more massive brown dwarfs \citep{sonora}. Older brown dwarfs therefore have smaller radii than younger, less-evolved brown dwarfs. \cite{wd1032} performed a kinematic analysis on \obj{} which determined it as a likely member of the thick disc within our Galaxy. As such, they derived an age estimate of 5--10~Gyr. Using the Sonora brown dwarf evolutionary models, a 70~M$_{\text{Jup}}$ brown dwarf at an age of 5~Gyr and an effective temperature of 1748~K should have a radius of $\sim$0.086~R$_{\odot}$. The radius of \objb{} measured from its eclipse is 0.1052~R$_{\odot}$, indicating that it is inflated. An inflated radius is also suggested for the non-eclipsing brown dwarf NLTT5306B, which \cite{rachael} propose can be explained by revising the age estimate to be significantly younger. If we consider a younger age, in order to have a radius of 0.1052~R$_{\text{Jup}}$, \obj{} would have to be only 500~Myr old. Since the cooling age of the white dwarf is 455~Myr, this would require a main sequence lifetime of only 45~Myr. This is unphysically short as a main sequence star would need to be at least 8.7~M$_{\odot}$ for its lifetime to be 45~Myr, and such a star would not evolve into a 0.4502~M$_{\odot}$ white dwarf. Therefore, we suggest that the radius of 0.1052~R$_{\text{Jup}}$ is larger than models would predict due to the constant irradiation slowing down the contraction of the brown dwarf. 

We note that of the irradiated brown dwarf companions to white dwarfs presented in Figure \ref{fig:HJWDBD}, only the two least irradiated systems, NLTT5306B and \objb{}, show evidence of inflation. As the white dwarf hosts of these brown dwarfs have lower effective temperatures, they have had more time to cool and pump heat into the interior of the brown dwarfs, which could slow the contraction and cause inflated radii \citep{casewell2020}. The longer wavelength irradiation from these white dwarfs is more easily absorbed by deeper layers of the atmosphere. Additionally, ultraviolet and shorter wavelength incident flux are more susceptible to scattering in the brown dwarf atmosphere, making it more difficult to reach the interior \citep{scattering}, and the brown dwarfs with higher levels of irradiation may have had more of their upper atmosphere dissociated by the ultraviolet heating.

It is possible that \objb{} migrated inwards towards the end of the main sequence lifetime of the white dwarf, and energy dissipation from the brown dwarf re-inflated it and set the evolution time to zero \citep{reinflate}. However, this is unlikely due to cooling age of the white dwarf. Another potential scenario for the inflation of \objb{} is that it is a post-bounce cataclysmic variable, where the orbital period has decreased and then increased again during evolution and mass transfer (e.g. \citealt{bounceCV}). The loss of angular momentum occurs on a comparable timescale to the thermal timescale of the donor star. This perturbs the donor star from its thermal equilibrium, causing it to have a slightly inflated radius compared to an isolated star \citep{CVML}. However, it is most likely that the constant irradiation from the white dwarf is slowing the contraction of the brown dwarf as it evolves, causing it to appear inflated. \objb{} is the only eclipsing white dwarf--brown dwarf binary which shows evidence that the brown dwarf is inflated.

\subsection{Potential Cataclysmic Variable Evolution}

We consider a potential evolution scenario for \obj{} if it is a detached cataclysmic variable. We calculate the Roche Lobe filling factor of \objb{} as 0.474, leading to the stellar surface of the brown dwarf having a maximum deformation from a perfect sphere of 3.5\%. The ellipsoidal variation is equal to $\sim$7\% of the total flux from the secondary, which corresponds to a 0.5\% ellipsoidal variability in the total flux of the system, which is negligible compared to the $\sim$8\% variation in flux which is introduced by irradiation and reflection.

\cite{Schreiber2023} proposed a mechanism where short period cataclysmic variables can detach due to the emergence of a magnetic field on the white dwarf, which transfers angular momentum from the spin of the white dwarf into the orbit and increases the orbital radius. Simulations of this evolutionary scenario were performed using the MESA code (\citealt{paxton11, paxton13, paxton15, paxton18, paxton19, jermyn23}, r24.03.1). The initial conditions are a detached post-common-envelope binary with a secondary mass of 0.65~M$_{\odot}$ and an initial period of 0.5~days. Evolution until the end of the cataclysmic variable phase assumed angular momentum loss using the same prescription as \citep{knigge11} which is tuned to reproduce the observed masses and radii of cataclysmic variable donor stars. Once the donor star reaches a mass of 69~M$_{\text{Jup}}$, we detach the binary by increasing the orbital period to 2.2~hours and follow the evolution of the secondary star alone.

It takes approximately 2~Gyr from the common envelope phase for the system to come into contact and reach a donor mass of69~M$_{\text{Jup}}$, at which point the newly detached secondary has a temperature of 2300~K and a radius of 0.11~R$_{\odot}$. It takes a further 0.2~Gyr for the secondary to shrink to the observed radius of WD1032+11B, at which point it has cooled to an effective temperature of 2100~K. During the cataclysmic variable phase the white dwarf would be heated by compressional heating to 10000--15000~K \citep{pala22}. The observed temperature of 9950~K is consistent with 0.2~Gyr of cooling from an initial temperature of 12500~K. The total age of the system would depend upon the initial orbital period of the binary, since a wider orbit would take longer to come into contact. We conclude that an evolutionary scenario whereby WD1032+011 is a detached cataclysmic variable is consistent with both the kinematic age and the observed radii and temperatures of the white dwarf and secondary star. The main objection to this explanation is that the spectrum of the white dwarf shows no sign of magnetism. Therefore if a strong field on the white dwarf is responsible for detaching the cataclysmic variable, the field must be short-lived.

\color{black}

\section{Conclusions}
\label{sec:conclusion}
We present Hubble Space Telescope Wide Field Camera 3 time-resolved spectrophotometry of the eclipsing white dwarf--brown dwarf binary \objb{}. We derive a broadband lightcurve which shows the primary eclipse, where the brown dwarf fully occults the white dwarf. Sub-band lightcurves do not show any wavelength-dependent changes in intensity, indicating that the irradiation equally penetrates the entire pressure range probed by our WFC3 spectra. We isolate the brown dwarf spectrum for different orbital phases including noon and midnight, and find that our dayside spectrum is on average 81\% brighter than our nightside spectrum. We calculate the brightness temperature across the entire spectral range, and find a 210~K difference between the dayside and nightside. Via comparison to field brown dwarfs, we identify the most likely spectral type of \objb{} as L1 pec. We use atmospheric retrievals to derive a dayside temperature of 1748$^{+66}_{-67}$~K and a nightside temperature of 1555$^{+76}_{-62}$~K. We do not recover clouds above the photosphere, and the dayside spectrum favours a cloud-free scenario, however the PICASO forward models fit the short wavelength data better when including clouds. We find that \objb{} is well-aligned in the mass--radius and temperature contrast--external irradiation flux parameter spaces of hot Jupiters and irradiated brown dwarfs. The brown dwarf radius is inflated compared to evolutionary models, and this inflation is likely driven by the irradiation from the white dwarf, which slows the brown dwarf's contraction. \objb{} is the only known inflated brown dwarf in an eclipsing white dwarf--brown dwarf binary, and upcoming JWST observations will further characterise the differences between its dayside and nightside atmospheres.

\section*{Acknowledgements}

J. R. French acknowledges support of a University of Leicester College of Science and Engineering PhD studentship. S. L. Casewell acknowledges the support of an STFC Ernest Rutherford Fellowship (ST/R003726/1). 
For the purpose of open access, the author has applied a Creative Commons Attribution (CC BY) licence to the Author Accepted Manuscript version arising from this submission.
Support for Programs number HST-GO-15947, number HST-AR-16142, and number HST-GO-16754 was provided by NASA through a grant from the Space Telescope Science Institute, which is operated by the Association of Universities for Research in Astronomy, Incorporated, under NASA contract NAS5-26555. HST data presented in this paper were obtained from the Mikulski Archive for Space Telescopes (MAST) at the Space Telescope Science Institute. This research has made use of the NASA Exoplanet Archive, which is operated by the California Institute of Technology, under contract with the National Aeronautics and Space Administration under the Exoplanet Exploration Program.
The authors would like to thank the anonymous reviewer for their helpful feedback.
%%%%%%%%%%%%%%%%%%%%%%%%%%%%%%%%%%%%%%%%%%%%%%%%%%
\section*{Data Availability}

The Hubble WFC3 data is available via the Mikulski Archive for Space Telescopes portal at \url{https://mast.stsci.edu/portal/Mashup/Clients/Mast/Portal.html}.



\bibliographystyle{mnras}
\bibliography{bib} % if your bibtex file is called example.bib

%\begin{thebibliography}{99}
%\end{thebibliography}

%%%%%%%%%%%%%%%%%%%%%%%%%%%%%%%%%%%%%%%%%%%%%%%%%%

%%%%%%%%%%%%%%%%% APPENDICES %%%%%%%%%%%%%%%%%%%%%

%%%%%%%%%%%%%%%%%%%%%%%%%%%%%%%%%%%%%%%%%%%%%%%%%%

\appendix
\section{Corner Plot of Broadband Lightcurve MCMC}

\begin{figure*}
    \centering
    \includegraphics[width=1.2\columnwidth]{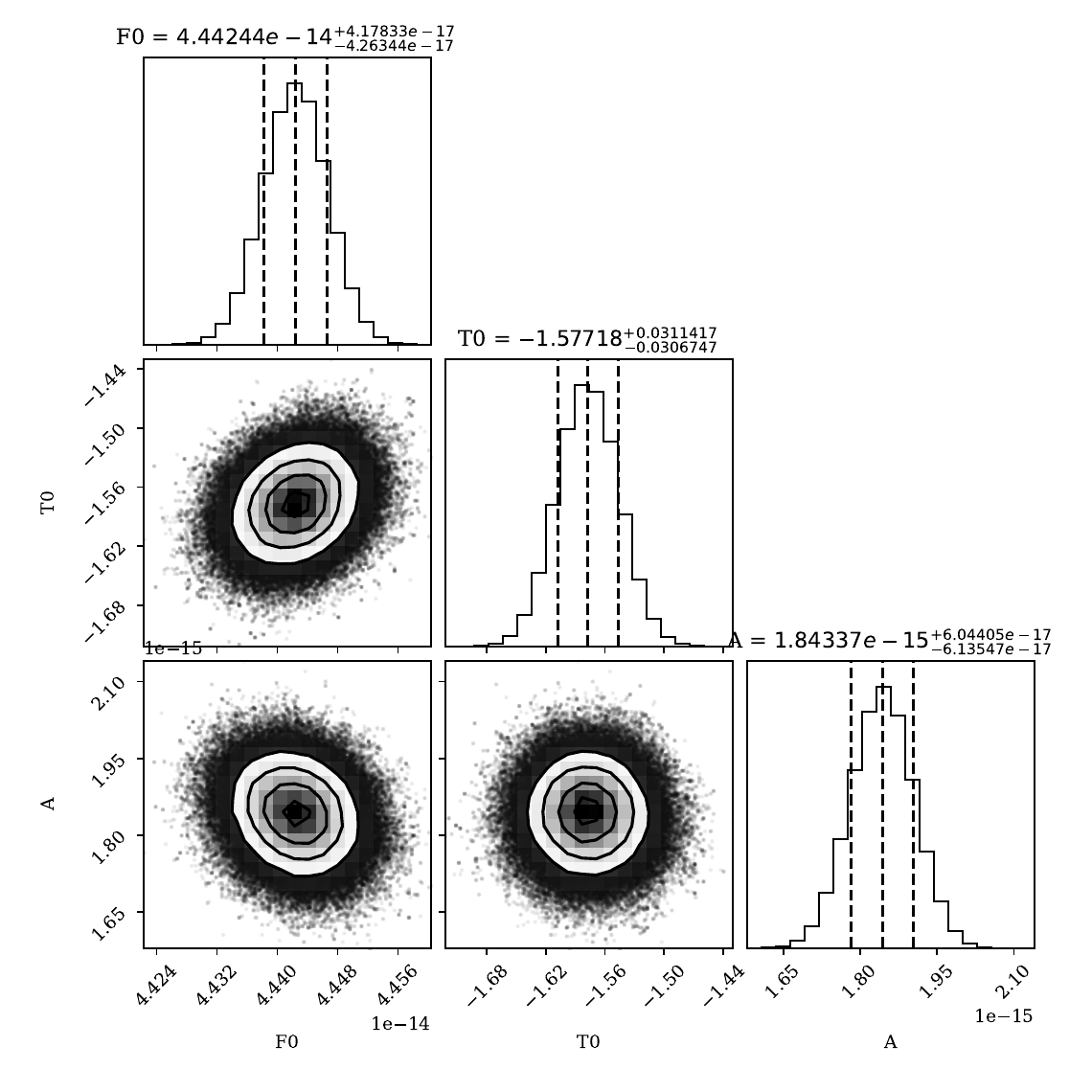}
    \caption{Corner plot showing the posterior distribution of the lightcurve model parameters for our MCMC fitting.}
    \label{fig:corner}
\end{figure*}

% Don't change these lines
\bsp	% typesetting comment
\label{lastpage}
\end{document}